\documentclass[12pt]{article}
\usepackage[a4paper, bindingoffset=0cm, hmargin={2.5cm, 2cm},vmargin={2.5cm, 2.5cm}]{geometry}

\usepackage{relsize}
\usepackage{graphicx}
\usepackage[english]{babel}
\usepackage[utf8]{inputenc}
\usepackage{comment}

\usepackage[T1]{fontenc}
\usepackage{lmodern}
\usepackage[bookmarks]{hyperref}
\usepackage{multirow}
\usepackage{float}

\usepackage{amsmath}
\usepackage{amssymb}
\usepackage{amsthm}
\usepackage{amsfonts}
\usepackage{mathtools}
\usepackage{dsfont}
\usepackage{braket}

\usepackage{pgfplots}
\usepackage{fancyhdr}
\usepackage{appendix}
\usepackage{bbold}
\usepackage{tablefootnote}
\usepackage{float}
\usepackage{amsmath,bm}
\usepackage{enumerate}
\usepackage{pgfplots}
\usepackage{epstopdf}
\usepgfplotslibrary{fillbetween}
\usetikzlibrary{patterns}
\pgfplotsset{compat=1.12}
\usepackage{mathtools}
\usepackage{caption}
\usepackage{subcaption}
\usepackage{cite}
\usepackage{stackrel}
\usepackage{bm}

\usepackage{color}
\theoremstyle{remark}

\usepackage{eso-pic}

\newcommand{\ab}[2]{\langle #1\,#2\rangle}

\newcommand{\req}[1]{(\ref{#1})}

\def\bet{\beta}
\def\fc#1#2{\frac{#1}{#2}}

\newcommand{\nwc}{\newcommand}
\nwc{\ba}  {\begin{array}}
\nwc{\ea}  {\end{array}}
\nwc{\bdm} {\begin{displaymath}}
\nwc{\edm} {\end{displaymath}}

\nwc{\bea} {\begin{equation}\ba{lcl}}
\nwc{\eea} {\ea\end{equation}}

\nwc{\be} {\begin{equation}}
\nwc{\ee} {\end{equation}}

\nwc{\bda} {\bdm\ba{lcl}}
\nwc{\eda} {\ea\edm}

\nwc{\bc}  {\begin{center}}
\nwc{\ec}  {\end{center}}

\nwc{\ds}  {\displaystyle}

\nwc{\nn} {\nonumber}
\nwc{\nnn} {\nonumber \vspace{.2cm} \\ }
\nwc{\ra}{\rightarrow}
\nwc{\lra}{\longrightarrow}
\def\lf{\left}\def\ri{\right}

\nwc{\p} {\partial}

\def\ap{\alpha'}

\def\eps{\epsilon}

\def\al{\alpha}
\def\bet{\beta}

\def\Sc{{\cal S}}

\def\eps{\epsilon}
\def\al{\alpha}

\def\bet{\beta}\def\bet{\beta}
\numberwithin{equation}{section}

\begin{document}

\thispagestyle{empty}
	\begin{center}

		{\bf\LARGE\sc Subleading Collinear Limits  of \\[4mm] Yang--Mills Amplitudes from Gravity}	
        \vspace{1.75cm}
		
		{{Jin Dong\ and\ Stephan Stieberger}
			
			\vspace{0.75cm}

			{\it\small
				Max-Planck Institut f\"ur Physik, Werner--Heisenberg--Institut,\\[2mm] Boltzmannstr. 8, 85748 Garching bei M\"unchen, Germany \\
			}
			
		}
		
	\end{center}
	\vspace{1.5cm}
	
	\begin{center}
		{\bf Abstract}\\
		\end{center}

		\noindent

We show that the subleading collinear sector of Yang--Mills (YM)
amplitudes  is
controlled by ordinary Einstein--Yang--Mills (EYM) amplitudes and
their higher-derivative corrections. The complete set of strict
subleading collinear limits associated with an equal-helicity collinear pair can be extracted
from open--closed string disk amplitudes.
The latter generate $\frac12 (n-3)!$ BCJ-like relations with non-linear kinematic
coefficients, reducing the $(n-3)!$ collinear data to a basis of dimension
$\frac12 (n-3)!$. This reduced basis can be represented by gravitational amplitudes
in EYM theories and by their higher-derivative open-string
corrections. At multiplicity $n$ this gravitational basis consists of the
$(n-4)!$ independent EYM subamplitudes together with
$
\frac12 (n-5)(n-4)!
$
higher-order corrections involving one graviton and $n-2$ gluons. The resulting
decomposition is governed by unsigned Stirling numbers of the first kind:
the ordinary EYM amplitudes correspond to the sector
$\left[{n-3\atop 1}\right]$, the higher-order BCJ-like relations to even
sectors $\left[{n-3\atop 2j}\right]$, and the higher-derivative EYM corrections
to odd sectors $\left[{n-3\atop 2j+1}\right]$ with $j\geq 1$.

	\vspace{1cm}
	\begin{flushright}
		{MPP-2026-144}
	\end{flushright}

\newpage

\tableofcontents

\newpage

\section{Introduction}

One of the most striking properties of string theory is the manifestation of gauge-gravity relations.
The prototypical example is provided by the Kawai--Lewellen--Tye relations, which express closed-string tree amplitudes on the sphere as bilinear combinations of open-string amplitudes \cite{KLT}. In the field-theory limit these relations become gravity/Yang--Mills relations and provide one of the conceptual origins of color-kinematics duality and the double-copy structure of gravitational amplitudes.

Color-ordered Yang--Mills $n$--point tree amplitudes are themselves subject to two
nested families of linear relations.  The Kleiss--Kuijf (KK) relations
reduce the set of $n!$ color orderings to a basis of $(n-2)!$ partial
amplitudes~\cite{KK}, while the Bern--Carrasco--Johansson (BCJ)
relations further reduce this basis to $(n-3)!$ independent
amplitudes~\cite{BCJ}.  The latter relations are closely tied to
color--kinematics duality and hence to the double-copy construction.
We shall use the abbreviations KK and BCJ throughout this work; for a
comprehensive review of color--kinematics duality and its applications,
see Ref.~\cite{Bern:2019prr}.

A complementary realization of gauge-gravity relations arises for mixed open-closed string amplitudes on the world--sheet disk. In this case, monodromy relations allow one to rewrite disk amplitudes with closed-string insertions in terms of purely open-string amplitudes with additional punctures \cite{Stieberger:0907}. This viewpoint was extended and applied in Refs. \cite{Stieberger:2014hba,Stieberger:2015kia_II,Stieberger:2015kia_PLB,Stieberger:2015kia,Stieberger:2016lng} to show that Einstein--Yang--Mills (EYM) amplitudes can be obtained from such open-closed disk amplitudes and, equivalently, that a graviton may be represented by a pair of collinear gauge bosons. Their work also revealed that subleading terms in the collinear expansion of tree--level Yang--Mills amplitudes are naturally constrained by EYM amplitudes.

The universal factorization of gauge-theory amplitudes in collinear limits is one of the most fundamental manifestations of locality and gauge invariance. While the leading collinear behaviour is completely universal, considerably less is known about the algebraic organization of strict subleading collinear data.
In this work we show that this sector possesses a remarkably rigid gravitational organization.

We extend the open-closed string perspective to the complete set of strict subleading collinear limits of (tree--level) gauge amplitudes. 
The relevant disk amplitudes generate not only the leading EYM relations but also higher-order BCJ-like relations with non-linear kinematic coefficients, together with higher-derivative corrections to EYM amplitudes. We show that the strict subleading collinear sector of Yang–Mills tree amplitudes admits a natural decomposition into gravitational building blocks. Starting from open–closed string disk amplitudes, we derive a hierarchy of BCJ-like relations which reduce the \((n-3)!\) independent Yang–Mills subleading collinear limits to a basis of dimension
$\frac12 (n-3)!$. The remaining independent data are completely organized by Einstein–Yang–Mills amplitudes and their higher-derivative open-string corrections involving one graviton and \(n-2\) gluons.
The main result of this work is the identification of a gravitational basis whose dimensions are given by the series of unsigned Stirling numbers of the first kind 
\be\label{Reihe}
\left[\!\begin{matrix}n-3\\1\end{matrix}\!\right],
\left[\!\begin{matrix}n-3\\3\end{matrix}\!\right],
\left[\!\begin{matrix}n-3\\5\end{matrix}\!\right],\ldots\ ,
\ee
while the complementary even Stirling sectors correspond to homogeneous BCJ-like constraints. The ordinary Einstein–Yang–Mills amplitudes span the sector $\left[\!\begin{matrix}n-3\\1\end{matrix}\!\right]=(n-4)!$, while successive higher-derivative Einstein–Yang–Mills sectors are described by the odd Stirling sectors $\left[\!\begin{matrix}n-3\\2j+1\end{matrix}\!\right]$ and the complementary even Stirling sectors $\left[\!\begin{matrix}n-3\\2j\end{matrix}\!\right]$ are exhausted by higher-order BCJ-like relations.
 Unsigned Stirling numbers already appear in Mafra's organization of
Kleiss--Kuijf--like relations among motivic components of pure-open superstring disk
amplitudes~\cite{Mafra:2021wok}.  However, the present result concerns a different vector
space and gives these numbers a different physical role.  After applying
the mixed open--closed monodromy projection, extracting the strict
subleading collinear coefficient and reducing to a BCJ basis, the even
Stirling blocks become homogeneous constraints, whereas the odd blocks
provide EYM and higher-derivative EYM coordinates on their common
solution space.  We verify the resulting complementary full-rank system and the  complete decomposition  explicitly through nine external particles  and conjecture its validity at arbitrary
multiplicity.
Hence our results reveal an unexpected combinatorial organization of subleading collinear Yang–Mills amplitudes arising  from the gravitational open–closed collinear sector. 

The observed Stirling ranks also have a suggestive mathematical
counterpart.  The Poincaré polynomial of the complement of the
type-\(A_{n-4}\) braid arrangement, equivalently the ordered
configuration space \(\operatorname{Conf}_{n-3}(\mathbb{C})\), encodes
the graded dimensions \cite{OrlikSolomon:1980}
\[
\dim H^{\,n-3-k}\!
\left(\operatorname{Conf}_{n-3}(\mathbb{C})\right)
=\left[{n-3\atop k}\right]\ .
\]
The Lehrer--Solomon decomposition organizes these cohomology spaces
by conjugacy classes of permutations with \(k\) cycles
\cite{LehrerSolomon:1986}, while Douglass, Pfeiffer and R\"ohrle
established a representation-theoretic relation between the
corresponding Orlik--Solomon components and the idempotent
decomposition associated with Solomon's descent algebra
\cite{DouglassPfeifferRoehrle}. 
Moreover, the factorization
\[
\operatorname{Conf}_{n-3}(\mathbb C)
\simeq
\mathbb C\times\mathbb C^\times\times\mathcal M_{0,n-2}
\]
relates this braid-arrangement geometry to the genus-zero moduli space
associated with the \(n-2\) original open-string insertions, up to an
affine-frame factor.
This provides a natural algebraic
counterpart to the associated graded dimensions of our Stirling
filtration.  However, at present no intertwining map between
these mathematical modules and the kinematically weighted row spaces
\(W_n^{(k)}\) is known.

This structure fits naturally into the broader gauge--gravity incarnation. While the original KLT relations concern tree-level closed-string amplitudes, subsequent developments, including  one-loop KLT-type relations   \cite{Stieberger:2021daa,Stieberger:2022lss,Stieberger:2023nol}, indicate that the open/closed organization of string amplitudes persists beyond tree level in a suitably generalized form. The collinear construction studied here may therefore be viewed as a disk-level, mixed open-closed analogue of this general gauge-gravity correspondence, refined to subleading order in the collinear expansion.

We further show that the hierarchy extends componentwise to hard
matter states in ${\cal N}=4$ super-Yang--Mills theory and study its
descendants under amplitude transmutation and dimensional reduction.
As an application, we recover the absence of a strict
$\epsilon^0$ term in adjacent collinear limits of NLSM amplitudes.

An independent field--theory analysis of subleading collinear limits was
undertaken in ~\cite{Nandan:2016CHY} using the
Cachazo--He--Yuan (CHY) representation of scattering
amplitudes \cite{Cachazo:2013hca,Cachazo:2013iea,Cachazo:2013gna}.  For two adjacent gluons of equal
helicity, they expressed the strict subleading collinear contribution as
an $(n-1)$-particle CHY integral weighted by a universal collinear
kernel and recovered the  EYM relations of \cite{Stieberger:2015kia_PLB} directly from
the scattering equations.  The kernel, however, depends on the
punctures and momenta of all hard particles and cannot be taken outside
the CHY integral.  Consequently, this representation does not by itself
provide a closed amplitude-level reconstruction of the individual
strict subleading collinear coefficients.  Among several other aspects the present work addresses
 this remaining reconstruction problem by combining the
open--closed disk relations, their higher-order monodromy constraints
and higher-derivative EYM amplitudes in a BCJ basis.

An additional motivation for the present construction comes from celestial holography, where collinear limits of momentum-space amplitudes determine operator-product expansions on the celestial sphere \cite{Fan:2019emx,Fotopoulos:2019vac}.  Whereas the singular celestial OPE is known rather generally \cite{PateRaclariuStromingerYuan2019,Fotopoulos:2019vac,Fotopoulos:2020bqj}, the organization of its regular terms beyond the MHV sector is considerably less transparent and is not manifestly local in the celestial coordinates \cite{PateRaclariuStromingerYuan2019,AdamoBuCasaliSharma2021,AdamoBuCasaliSharma2022,RenSchreiberSharmaWang2023}.  The gravitational reconstruction found here suggests a different organization: the EYM amplitudes and their higher-derivative corrections provide candidate matrix elements of local gravitational blocks associated with the fused momentum \(P\).  Establishing this interpretation requires extending the present symmetric momentum partition to arbitrary \(x\) and performing the corresponding Mellin transforms.  We outline this prospective application in section~\ref{celestial}.

The symmetric momentum partition of the two collinear momenta (\(x=\tfrac12\))  used throughout this
work is distinguished not only by the left--right symmetry of the
open--closed disk amplitude.  A general field-theory analysis of the
pole-subtracted collinear limit shows that the dependence of the
strict finite coefficient on the auxiliary reference spinor \(r\),
which specifies the direction of approach to the collinear surface,
is proportional to \(1-2x\) \cite{Jin}.  Consequently, at
\(x=\tfrac12\) every subleading coefficient \(C_\rho^{(0)}\) is individually
independent of \(r\).  The reconstruction problem considered below
therefore concerns intrinsic data on the collinear surface rather than
data associated with a particular off-collinear continuation. 

This work  is organized as follows. In section 2 we review the strict subleading collinear limit of Yang–Mills amplitudes and its relation to mixed open–closed disk amplitudes, and formulate the counting problem which motivates the subsequent construction. In section 3 we define the strict collinear projection, discuss the pole-mixing term associated with an adjacent BCJ basis, and state the all-multiplicity hierarchy of EYM data and homogeneous BCJ-like constraints. Section 4 derives this hierarchy explicitly from the open–closed string relations at five, six and seven points and then exhibits its generic structure. In section 5 we assemble these relations into the complete reconstruction system, establish the Stirling-number rank pattern through nine points, and discuss its relation to known Stirling counting, the Solomon descent algebra and the Orlik–Solomon structure of the type-\(A\) braid arrangement. Section 6 extends the construction to other states in \(\mathcal N=4\) super–Yang–Mills theory and to effective field theories obtained by transmutation and dimensional reduction. We conclude in section 7 with a discussion of the resulting gravitational basis, the extension of our approach to mixed-helicity collinear pairs, and how EYM amplitudes and their higher-derivative corrections may possibly furnish a gravitationally organized basis of candidate local blocks for regular celestial OPEs beyond the MHV sector. Appendix A provides a local world-sheet interpretation of the pole-mixing term.
\section{From Yang--Mills collinear limits to open--closed disk amplitudes}

Throughout sections~2--\ref{sec:universality}, we restrict the adjacent collinear pair to equal helicities, \((h_{n-1},h_n)=(+,+)\) or \((-,-)\). The extension to mixed-helicity pairs is discussed in section~7.2.

\subsection{Subleading collinear limits in Yang--Mills theory}

We consider an \(n\)-point color-ordered tree--level YM amplitude in the adjacent collinear limit
\(p_{n-1}\parallel p_n\).
The collinear limit is defined as the kinematic configuration in which two massless 
particles $i=n-1$ and $j=n$ propagate with parallel momenta. Following 
\cite{Stieberger:2015kia_PLB}, we introduce light-like vectors $P$ and $r$ and 
parametrise the momentum spinors as:
\begin{equation} \label{eq: col kin}
\begin{aligned}
\lambda_{n-1} &= \lambda_P\cos\theta - \epsilon\,\lambda_r\sin\theta\,,&
\tilde\lambda_{n-1} &= \tilde\lambda_P\cos\theta - \epsilon\,\tilde\lambda_r\sin\theta\,,\\
\lambda_n &= \lambda_P\sin\theta + \epsilon\,\lambda_r\cos\theta\,,&
\tilde\lambda_n &= \tilde\lambda_P\sin\theta + \epsilon\,\tilde\lambda_r\cos\theta\,,
\end{aligned}
\ee
with $c=\cos\theta=\sqrt{x}$, $s=\sin\theta=\sqrt{1-x}$. The collinear limit 
is reached as $\epsilon\to 0$.  We distinguish the dimensionful
field-theory invariants from their dimensionless string counterparts by
\begin{equation}\label{Mandelstam}
 s_{ij}^{(0)}:=(p_i+p_j)^2=2p_i\!\cdot p_j,
 \qquad
 s_{ij}:=\alpha' s_{ij}^{(0)},
 \qquad
 s_{iP}^{(0)}:=2p_i\!\cdot P.
\end{equation}
The dimensionless invariants are expanded as (with $s_{ij}^{(0)}=\langle ij\rangle[ji]$):
\begin{equation}\label{MandelO}
\begin{aligned}
s_{i,n-1} &= \alpha' x\;s_{iP}^{(0)} +\alpha'\epsilon\ sc\left\{\;\langle i\; P\rangle[ir]+\langle i r\rangle[i\; P]\;\right\}+\ap^2\eps^2 s^2\ (s_{ir}^{(0)})^2\ ,\\
s_{i,n} &= \alpha'(1-x)\;s_{iP}^{(0)}-\alpha'\epsilon\ sc\left\{\;\langle i\; P\rangle[ir]+\langle i r\rangle[i\; P]\;\right\}+\ap^2\eps^2 c^2\;(s_{ir}^{(0)})^2\ ,\\
s_{n-1,n}&=\alpha'(p_{n-1}+p_n)^2
= \alpha'\epsilon^2\;\langle Pr\rangle\;[rP] \ .
\end{aligned}
\ee
Partial amplitudes with \emph{adjacent} $n-1,n$ contain simple poles $O(\epsilon^{-1})$ 
at leading order. The subleading, finite terms at $O(\epsilon^0)$ are the focus of 
this work.
We define
\be
A_{\rm YM}(1,\rho,n-1,n)
=
\frac{1}{\epsilon}\ C_\rho^{(-1)}
+\epsilon^0\ 
C_\rho^{(0)}
+
O(\epsilon)\ ,\label{ExpaAmp}
\ee
and the goal is to determine the corrections:
\be\label{CollExpa}
C_\rho^{(0)}:=A_{\rm YM}(1,\rho,n-1,n)_{\rm sub}\ .
\ee
Throughout this work, we restrict the two collinear gluons to equal helicities, cf. subsection \ref{MixedHelicity} for generalizing our approach.

In a BCJ basis there are \((n-3)!\) independent partial
amplitudes, and hence we expect \((n-3)!\) independent strict subleading coefficients
\(C_\rho\) at order \(\epsilon^0\).
For MHV  $n$-point amplitudes we determine the following closed form expression for the subleading correction
\begin{align}
A\!\left(1^-,2^-,3^+,\ldots,n^+\right)_{\rm sub}
&=-\frac{1}{x(1-x)}
\frac{(x-1)\,\ab{r}{(n-2)}\,\ab{1}{P} + x\,\ab{r}{1}\,\ab{(n-2)}{P}}
{\ab{r}{P}\,\ab{1}{P}\,\ab{(n-2)}{P}}\nonumber\\
&\times A\!\left(1^-\,2^-\,3^+,\ldots,P^+\right)\nonumber\\
&=\Big\{\frac{1}{x}\frac{\ab{r}{(n-2)}}
{\ab{r}{P}\ab{(n-2)}{P}}-\frac{1}{1-x}\frac{\ab{r}{1}}
{\ab{1}{P}\ab{r}{P}}\Big\}\nonumber\\
&\times A\!\left(1^-,2^-,3^+,\ldots,P^+\right)\ ,
\end{align}
with $x$ and reference spinor $r$. 

On the other hand, NMHV results are tedious. While expressions for six--point NMHV are given for non--adjacent collinear legs in \cite{Stieberger:2015kia_PLB} here we give a sample of adjacent collinear legs
\begin{align}
A\!\left(1^-,2^-,3^-,4^+,5^+,6^+\right)_{\rm sub}
&=\frac{1}{\text{s}^2 \langle 1\,P\rangle \langle 4\,P\rangle \langle r\,P\rangle}\ \Big\{
\frac{\text{s}^2 \langle 1\,P\rangle \langle r\,4\rangle-\text{c}^2 \langle 4\,P\rangle \langle r\,1\rangle}{\text{c}^2}\nonumber\\
&\quad-\frac{\text{c}^2 s_{12}[3\,2]\langle 1\,2\rangle
(\langle 4\,P\rangle\langle r\,3\rangle-\langle 3\,P\rangle\langle r\,4\rangle)}
{s_{34}\left(\text{c}^2s_{12}+\text{s}^2s_{34}\right)}\nonumber\\
&\quad+\frac{[3\,2]}{[2\,1][P\,1]}
\Big[-\frac{(s_{12}+s_{34})[P\,1]}{[4\,3]}\langle r\,P\rangle
+[2\,1]\langle 3\,4\rangle\langle r\,2\rangle \nonumber\\
&\quad+([2\,1]\langle 2\,3\rangle+[P\,1]\langle 3\,P\rangle)
\langle r\,4\rangle
-([2\,1]\langle 2\,4\rangle+[P\,1]\langle 4\,P\rangle)
\langle r\,3\rangle\Big]\Big\}\nonumber\\
&\quad\times A\!\left(1^-,2^-,3^-,4^+,P^+\right)\ .\label{NMHV}
\end{align}
Note that the same split factor holds for $A\!\left(1^-,2^-,3^+,4^-,5^+,6^+\right)_{\rm sub}$, $A\!\left(1^-,2^+,3^-,4^-,5^+,6^+\right)_{\rm sub}$ and $A\!\left(1^+,2^-,3^-,4^-,5^+,6^+\right)_{\rm sub}$.

Within the pole-subtraction prescription of Ref.~\cite{Jin}, the
general field-theory analysis of the adjacent collinear limit implies
the structural decomposition
\begin{equation}
 C_\rho^{(0)}(x;r)
 =
 C_{\rho,\mathrm{reg}}^{(0)}(x)
 +(1-2x)\,
 C_{\rho,\mathrm{pole}}^{(0)}(x;r),
 \label{reference-decomposition}
\end{equation}
where \(C_{\rho,\mathrm{reg}}^{(0)}(x)\) is obtained by restricting the
pole-free remainder of the amplitude to the collinear surface, while
\(C_{\rho,\mathrm{pole}}^{(0)}(x;r)\) contains the finite contribution
generated by the adjacent factorization pole.  The former is
independent of the direction $r$ of approach, and all dependence on the
auxiliary reference spinor \(r\) is confined to the latter
\cite{Jin}.  It follows that
\begin{equation}
 C_\rho^{(0)}\!\left(\frac12;r\right)
 =
 C_{\rho,\mathrm{reg}}^{(0)}\!\left(\frac12\right)
 =
 C_\rho^{(0)}\!\left(\frac12;r'\right)
 \label{reference-independence}
\end{equation}
for any admissible reference spinors \(r\) and \(r'\).  Thus the strict
subleading coefficient of each color ordering, and not merely a
particular linear combination of such coefficients, is independent
of the off-collinear continuation at the symmetric splitting.
E.g. for $x=\tfrac12$ (\ref{NMHV}) reduces to the $r$--independent expression:
\begin{align}
A\!\left(1^-,2^-,3^-,4^+,5^+,6^+\right)_{\rm sub}&=
\frac{2}{\langle1P\rangle\langle4P\rangle}\;\left(\langle14\rangle -\frac{\langle12\rangle\langle34\rangle[32]}{s_{12}+s_{34}}\right)
 A\!\left(1^-,2^-,3^-,4^+,P^+\right)\ .\label{NMHVa}
\end{align}
This provides an explicit non-MHV check of the general statement
\eqref{reference-independence}.

\subsection{Open--closed disk amplitudes and EYM amplitudes}

At tree level in string theory, scattering processes involving both open and closed 
strings are described by a disk world-sheet. The upper half-plane 
$\mathbf{H}_+ = \{z \in \mathbb{C}\,|\,\mathrm{Im}(z)\geq 0\}$ provides a convenient 
parametrisation. Then open string vertex operators are inserted at the boundary 
(real axis), while closed string operators are inserted in the bulk.

We consider disk amplitudes $A(1,2,\ldots,n-2;\,q_1,q_2)$ with $n-2$ open strings 
carrying momenta $p_i$ ($i=1,\ldots,n-2$) and one closed string whose left-- and 
right--moving momenta are $q_1$ and $q_2$ respectively. In the presence of D-branes, 
these are split as
\begin{equation}\label{Dmom}
q_1 = \tfrac{1}{2}q\,,\qquad q_2 = \tfrac{1}{2}Dq\,,
\end{equation}
with $D$ a matrix encoding the boundary conditions. Momentum conservation along 
the D-brane world-volume reads 
\be\label{Momcons}
\sum_{i=1}^{n-2}p_i + q^\parallel = 0\ ,
\ee
where 
$q^\parallel = q_1+q_2$.
In the collinear limit we consider the momenta 
\bea\label{leadingcoll}
p_{n-1}&=&q_1=xP\ ,\\
p_n&=&q_2=(1-x)P\ .
\eea

\def\floor#1{\lfloor#1\rfloor}
A central result of \cite{Stieberger:2015kia} is that 
the mixed disk amplitudes $A(1,2,\ldots,n-2;\,q_1,q_2)$ can be expressed as linear combinations of pure open string amplitudes. 
Our starting point is the expression 
\begin{align}
A(1,2,\dots,n{-}2;q_1,q_2)&=(-1)^n\ e^{-\pi i (s_{1,n}+s_{2,n-1})}\
\sum_{l=2}^{n-2}(-1)^l\ \sin(\pi s_{l,n-1})\ e^{\pi i (-1)^l\ s_{l,n-1}} \nonumber\\
&\times\sum_{\rho\in \{OP(\al,\beta^t),l\}}
e^{\pi i\sum\limits_{k=1}^{\floor{\frac{n-3}{2}}}\tau_{2k+1}(\rho)}\
\Sc(\rho)\ A(1,\rho,n-1,n)\ ,\label{FINAL}
\end{align}
constituting a sum of $2^{n-4}$ terms. We have introduced the $n$ open string momenta $k_i=p_i,\;i=1,\ldots,n-2$, $k_{n-1}=q_1$ and $k_n=q_2$ and the kinematic invariants $s_{ij} = 2\alpha' k_i\cdot k_j$. The second sum involves all  permutations $\rho$ comprising the element $l$ and the ordered set of permutations $OP(\al,\bet^t)$ of the merged sets:
\be\label{sets}
\al=\{2,\ldots,l-1\}\ \ \ ,\ \ \ \bet=\{l+1,\ldots,n-2\}\ .
\ee
This ordered set corresponds to all permutations of $\al\cup \bet^t$ which keep the order of elements
of $\al$ and $\bet^t$, respectively. Besides, $\bet^t$ denotes reversal of the elements in $\bet$.
Furthermore, in \req{FINAL} the following string kernel $\Sc(\rho)$ enters
\be\label{kernel}
\Sc(\rho)\equiv \Sc[ \rho(2,\ldots,n-2) \, ] = \prod_{i=2}^{n-2}\prod_{j=i+1}^{n-2}
\exp\lf\{  \pi i\;\Theta(\rho^{-1}(i)-\rho^{-1}(j)) \ s_{i,j} \ri\}\ ,
\ee
with $\Theta$ denoted the Heaviside step function.
Some other variants of string KLT kernels occur for pure closed string amplitudes in \cite{KLT}. Finally, we have:
\be\label{TAU}
\tau_i(\rho)=\begin{cases}
{\rm sign}(\rho^{-1}(i)-\rho^{-1}(i+1))\ (s_{i,n-1}+s_{i+1,n-1})\ ,&3\leq i\leq n-3\ ,\\
s_{n-2,n-1}\ ,&i=n\!-\!2\ .
\end{cases}
\ee
To illustrate \req{FINAL} for $n=5$ we have
\be\label{finalf}
A(1,2,3;q_1,q_2)=e^{-\pi i s_{24}}\ \lf[\ e^{-\pi i s_{51}}\ \sin(\pi s_{34})\ A(1,2,3,4,5)-
\sin(\pi s_{24})\ A(1,3,2,4,5)\ \ri]\ ,
\ee
while for $n=6$ \req{FINAL} yields:
\begin{align}
    A(1,2,3,4;q_1,q_2)&=e^{-\pi i s_{25}}\ \lf\{\ e^{-\pi i (s_{61}+s_{35})}\ \sin(\pi s_{45})\ A(1,2,3,4,5,6)\ri.\nonumber\\
&+\sin(\pi s_{25})\ A(1,4,3,2,5,6)-e^{\pi i (-s_{61}+s_{34}+s_{45})}\ \sin(\pi s_{35})\nonumber\\
&\lf.\times\lf[\ A(1,2,4,3,5,6)+e^{\pi i s_{24}}\ A(1,4,2,3,5,6)\ \ri]\ \ri\}.
\label{finals}
\end{align}

The full string corrections to \req{FINAL} are furnished by
inserting the compact expression\footnote{The ordering colons $:\ldots :$ are defined such that matrices with larger subscript multiply matrices with smaller subscript from the left, i.e.~$: \, M_{i} \ M_{j} \, : =  
\begin{cases}M_{i} \ M_j\ , & i \geq j\ ,\\ 
       M_{j} \ M_i\ , &              i<j\ .\end{cases}$
The generalization to iterated matrix products $: M_{i_1} M_{i_2} \ldots M_{i_p}:$ is straightforward.} \cite{Schlotterer:2012ny}
\be\label{EXPA}
\vec A=F\ \vec A_{\rm YM}\ \ \ ,\ \ \ F=P\ Q\ :e^{\sum\limits_{k\geq1}\zeta_{2k+1}M_{2k+1}}:\ ,
\ee
with the $(n-3)!$ dimensional  open string  basis vector $\vec A^t=\{A(1,\rho,n-1,n)\ |\ \rho\in S_{n-3}\}$ specified in \cite{Mafra:2011nv, Mafra:2011nw}. The $(n-3)!\times(n-3)!$ matrices $P,M$ can systematically be determined recursively from the Drinfeld associator \cite{Broedel:2013aza}

In the collinear (EYM) limit \req{leadingcoll}, at the leading order 
in $\alpha'$ we obtain Einstein--Yang--Mills amplitudes $A_{\rm EYM}$. Specifically, for a graviton 
with momentum $P$ and $n-2$ gluons, for $n=5$ the result   reads \cite{Stieberger:2014hba}
\be
A_{\rm EYM}(1,2,3;P)=\frac{\kappa}{g^2}\ x\; 
\lf[\ s_{3P}\ A(1,2,3,4,5)-s_{2P}\ A(1,3,2,4,5)\ \ri]\ ,
\label{eq:EYM}
\ee
while for $n=6$ we have \cite{Stieberger:2014hba}:
\begin{align}
    A_{\rm EYM}(1,2,3,4;P)&=\frac{\kappa}{g^2}\ x\;\lf\{\  s_{4P}\ A(1,2,3,4,5,6)+s_{2P}\ A(1,4,3,2,5,6)\ri.\nonumber\\
&\lf.-s_{3P}\ \lf[\ A(1,2,4,3,5,6)+ A(1,4,2,3,5,6)\ \ri]\ \ri\}\ .\label{eq:EYM6}
\end{align}
Moreover, in the  limit $x=\tfrac12$ the closed string momenta \req{Dmom} are identical for the left-- and right--mover and
the result \req{FINAL} must be real to all orders in $\ap$, i.e.:
\be\label{NYC}
\Im A(1,2,\dots,n{-}2;q_1,q_2)\;\Big|_{q_1 = q_2 = \tfrac{1}{2}P}=0\ .
\ee
The imaginary part
of \req{FINAL} appears at even orders in $\ap$. Its origin 
arises from the monodromy phases in \req{FINAL} accounting for the splitting of the complex closed string integration into two real open string integrals. The condition \req{NYC} describes the property that the closed-string
graviton state behaves left--right symmetrically on the disk
world-sheet.  The symmetric point is therefore characterized by two
independent properties:
\be\label{Situation}
x=\frac12
\quad\Longrightarrow\quad
\begin{cases}
q_1=q_2=\dfrac{P}{2}
\quad\Longrightarrow\quad
\displaystyle
\operatorname{Im}
A(1,\ldots,n-2;q_1,q_2)=0,
\\[3mm]
\displaystyle
C_\rho^{(0)}\!\left(\frac12;r\right)
=
C_\rho^{(0)}\!\left(\frac12;r'\right),
\qquad
\rho\in S_{n-3},
\end{cases}
\ee
for any admissible reference spinors \(r\) and \(r'\).  The first line
expresses the left--right symmetry of the open--closed disk amplitude,
whereas the second follows from the field-theory result
\eqref{reference-independence}.  Thus the symmetric open--closed
relations act on intrinsically defined strict collinear data.

The strict reality condition \req{NYC} can be refined as off--collinear reflection property in the following way. Complex conjugation of the underlying world--sheet disk integral \req{eq:mixed-disk-integral-main} of the amplitude  $A(1,2,\ldots,n-2;\,q_1,q_2)$ on a
real kinematic slice exchanges the bulk coordinates
\(z\leftrightarrow\bar z\) and therefore interchanges the left- and
right-moving data.  For a left--right symmetric closed-string state, this gives:
\begin{equation}
 A(1,\ldots,n-2;q_2,q_1)=A(1,\ldots,n-2;q_1,q_2)^* .
 \label{eq:left-right-reflection}
\end{equation}
At \(x=\tfrac12\), the collinear parametrization
\req{eq: col kin} obeys:
\begin{equation}
 q_1(-\epsilon)=q_2(\epsilon),
 \qquad
 q_2(-\epsilon)=q_1(\epsilon)\ .
 \label{eq:epsilon-exchange}
\end{equation}
For an exactly momentum-conserving continuation \req{Momcons} whose
\(O(\epsilon^2)\) completion is even under
\(\epsilon\to-\epsilon\), the two complete kinematic configurations
differ only by the exchange \(q_1\leftrightarrow q_2\).
Equations~\req{eq:left-right-reflection} and
\req{eq:epsilon-exchange} therefore imply:
\begin{equation}
 A(1,\ldots,n-2;q_1(-\epsilon),q_2(-\epsilon))
 =
 A(1,\ldots,n-2;q_1(\epsilon),q_2(\epsilon))^* .
 \label{eq:epsilon-reflection}
\end{equation}
In particular, the coefficient of \(\epsilon^0\) in the complete
mixed disk amplitude is real, in agreement with and refining the
strict reality condition \req{NYC}. Moreover, after representing the amplitude $A$ as Laurent series in $\epsilon$ and applying the relation \req {eq:epsilon-reflection} the coefficient of \(\epsilon^{m}\) has to be  real for even $m$ and is imaginary for odd $m$, including the leading divergence at $m=-1$.

\subsection{The counting problem: missing equations}

In the EYM amplitudes \req{eq:EYM}, \req{eq:EYM6} and in the subsequent higher multiplicity $n$ expressions only strict subleading collinear SYM amplitudes are contributing, since their leading contributions cancel as a result of BCJ relations. The strict subleading collinear sector of an \(n\)-point YM amplitude contains
\((n-3)!\) independent coefficients, one for each element of a BCJ basis \cite{BCJ}.  The
ordinary one-graviton EYM amplitudes obtained from the leading real part of the
open--closed disk amplitude \req{FINAL} do not provide enough independent data to determine all
of them.  Indeed, the color-ordered EYM amplitudes with \(n-2\) gluons and one
graviton obey KK/U(1)-decoupling relations \cite{Stieberger:2016lng}, and therefore span only a space of
dimension $(n-4)!$.

This mismatch is already visible at six points. The six strict subleading YM
coefficients (four of them entering \req{eq:EYM6}) are constrained by only two independent EYM amplitudes e.g. $A_{\rm EYM}(1,2,3,4;P)$ and $ A_{\rm EYM}(1,3,2,4;P)$.  The missing information must therefore come either from additional (BCJ--like) relations among the
subleading YM data or from higher string corrections to the EYM sector.

The examples studied in section \ref{Results} suggest a natural organization of the missing information.
The monodromy expansion of the open--closed disk amplitude \req{FINAL} generates homogeneous
BCJ-like relations with higher-degree kinematic kernels, while the complementary
data are represented by ordinary and higher-derivative EYM amplitudes.  The resulting
pattern is governed by unsigned Stirling numbers of the first kind.
The missing information is supplied by higher-order BCJ-like relations \req{NYC} (at orders $\ap^{2n}$) and by
higher-derivative EYM corrections $A_{\rm EYM}^{F^{2n}}$ (at orders $\ap^{2n-1}$) to be specified below.
\section{Collinear projection and all--multiplicity hierarchy}
\label{strictprojection}

After reducing all open-string amplitudes to the Yang--Mills basis by
means of \req{EXPA}, we may express \req{FINAL} as 
\begin{equation}
 A(1,\ldots,n-2;q_1(\epsilon),q_2(\epsilon))
 =\sum_{\rho\in S_{n-3}}
 c_\rho(\epsilon,\alpha')\,
 A_{{\rm YM},\rho}(\epsilon),
 \qquad
 A_{{\rm YM},\rho}(\epsilon)
 :=
 A_{\rm YM}(1,\rho,n-1,n),
 \label{eq:effective-open-closed-relation}
\end{equation}
where \(c_\rho\) is the complete effective coefficient containing both
the monodromy kernel \req{kernel} and the open-string period matrix $F$ 
defined in \req{EXPA}. In particular, expanding  $c_\rho(\epsilon,\alpha')$ w.r.t. small $\alpha'$ starts at the linear order.
We use the adjacent BCJ basis throughout because its coordinates are precisely the strict subleading collinear coefficients that we wish to reconstruct.

\subsection{Exact momentum conservation and the order of limits}
\label{sec:exact-collinear-expansion}

The important point is that all relations below, including
\req{eq:set1} and the explicit results of section~\ref{Results}, are
extracted from the {\it exact} open--closed disk amplitude expressions 
\req{FINAL} and \req{eq:effective-open-closed-relation}.  We evaluate
these amplitudes on a one-parameter family of {\it exactly}
momentum-conserving \(n\)-particle configurations satisfying
\req{Momcons}.  The endpoint at \(\epsilon=0\) obeys the strict
collinear replacement \req{leadingcoll}, while the two effective
open-string momenta are continued away from the collinear surface
according to \req{eq: col kin},
\[
q_1(\epsilon):=p_{n-1}(\epsilon),
\qquad
q_2(\epsilon):=p_n(\epsilon).
\]
Since this parametrization gives
\[
q_1(\epsilon)+q_2(\epsilon)=P+\epsilon^2r,
\]
exact momentum conservation at finite \(\epsilon\) may be maintained
either by treating \(P_{\rm tot}:=q_1+q_2\) as the exact parent momentum
in the remaining hard kinematics, or equivalently by an
\(O(\epsilon^2)\) deformation of spectator momenta.  At the symmetric
partition \(x=\tfrac12\), this completion can be chosen even under
\(\epsilon\to-\epsilon\).

Only after evaluating the exact disk relations on this
momentum-conserving family do we expand in the collinear parameter
\(\epsilon\).  This expansion must be distinguished from the
independent expansion in the string parameter \(\alpha'\).  At each
fixed order in \(\alpha'\), the effective coefficients
\(c_\rho(\epsilon,\alpha')\) admit a Taylor expansion in \(\epsilon\),
whereas the Yang--Mills amplitudes and hence the complete mixed disk
amplitude admit Laurent expansions.

\subsection{Pole mixing and the strict projection}

Near the adjacent collinear channel, these $\epsilon$--expansions take the form:
\begin{equation}
 \begin{aligned}
 c_\rho(\epsilon,\alpha')
 &=
 c_{\rho,0}(\alpha')
 +\epsilon\,c_{\rho,1}(\alpha')
 +O(\epsilon^2),\\
 A_{{\rm YM},\rho}(\epsilon)
 &=
 \frac{C_\rho^{(-1)}}{\epsilon}
 +C_\rho^{(0)}
 +O(\epsilon).
 \end{aligned}
 \label{cExp}
\end{equation}
Consequently, the coefficient of \(\epsilon^0\) in the complete mixed
disk amplitude contains two contributions, 
\begin{equation}
 \left.
 A(1,\ldots,n-2;q_1(\epsilon),q_2(\epsilon))
 \right|_{\epsilon^0}
 =
 \sum_{\rho}
 \left(
 c_{\rho,0}C_\rho^{(0)}
 +
 c_{\rho,1}C_\rho^{(-1)}
 \right).
 \label{eq:full-finite-with-mixing}
\end{equation}
The second term\footnote{The appearance of the pole-mixing term is tied to our use of an adjacent BCJ basis $\{A_{{\rm YM},\rho}|\rho\in S_{n-3}\}$. One could instead choose a BCJ basis in which the two collinear legs are non-adjacent. The corresponding basis amplitudes are regular in the $p_{n-1}\parallel p_n$ channel, and hence no term ${\cal B}$ of the form $c_{\rho,1}C_\rho^{(-1)}$ arises directly. This does not, however, eliminate the mixing relevant for the present reconstruction problem. The BCJ transformation from such a non-adjacent basis to the adjacent basis becomes singular on the collinear locus, involving inverse powers of $s_{n-1,n}={\cal O}(\epsilon^2)$. Consequently, the finite adjacent coefficients $C_\rho^{(0)}$ receive contributions from different orders in the collinear expansion of the non-adjacent basis amplitudes. The strict projection therefore isolates this mixing explicitly rather than shifting it into a singular change of basis.}, 
\begin{equation}
 {\cal B}:=
 \sum_\rho c_{\rho,1}\ C_\rho^{(-1)},
 \label{eq:pole-mixing}
\end{equation}
is the first off-collinear variation of the effective string coefficient
multiplying the already known universal leading Yang--Mills pole.  A concise world-sheet
interpretation of this pole-mixing term ${\cal B}$ is given in
appendix~\ref{app:pole-mixing}. It does
not contain new strict subleading Yang--Mills data and depends in general
on the chosen continuation away from the collinear surface.
This statement should not be confused with a reality property of the
individual leading coefficients \(C_\rho^{(-1)}\).  At the leading
field-theory order relevant to ordinary EYM \cite{Stieberger:2015kia_PLB}, the (real) coefficient of the
\(1/\epsilon\) pole in the complete open--closed combination vanishes by
the homogeneous BCJ relation \cite{BCJ}:
\begin{equation}
 \sum_\rho c_{\rho,0}\big|_{\alpha'}\ C_\rho^{(-1)}=0 \ .
 \label{eq:leading-pole-bcj-cancellation}
\end{equation}
Here \(c_{\rho,0}|_{\alpha'}\) denotes the leading field-theory component
of the effective coefficient \(c_{\rho,0}\), which starts linearly in $\alpha'$. Thus the leading field-theory contribution starts at finite collinear order. However, at higher \(\alpha'\) orders, a leading 
\(\epsilon^{-1}\) divergence may be present and, whenever non-vanishing, is purely imaginary according to~\eqref{eq:epsilon-reflection}. The world-sheet origin of this divergence is discussed in Appendix~\ref{app:pole-mixing}. In what follows, we focus primarily on the finite $\epsilon^{0}$ contribution.

The EYM relations are formulated directly on the collinear surface, with
\(q_1=xP\) and \(q_2=(1-x)P\).  We therefore freeze the complete effective
coefficient on this surface before extracting the finite Yang--Mills
coefficient and define:
\begin{equation}
 \left.
 A(1,\ldots,n-2;q_1,q_2)
 \right|_{\rm strict,\,\epsilon^0}
 :=
 \sum_{\rho\in S_{n-3}}
 c_{\rho,0}\ C_\rho^{(0)}\, .
 \label{eq:strict-projection-main}
\end{equation}
This strict projection omits ${\cal B}$, while retaining every
strict subleading coefficient \(C_\rho^{(0)}\).  The same prescription is
used in section \ref{Results} for the ordinary and higher-derivative EYM sectors.  The term ``strict'' does not refer to an additional kinematic limit.
It denotes the projection of the finite coefficient onto the value of
the effective kernel on the collinear surface.  The complete
\(\epsilon^0\) coefficient of the disk amplitude additionally contains
the pole-mixing contribution \(\mathcal B\).

At \(x=\tfrac12\), each \(C_\rho^{(0)}\) is independently intrinsic by
\req{reference-independence}, so that the projected relations do not depend
on the auxiliary reference spinor \(r\).  
The strict projection does not replace the exact momentum-conserving
procedure described above.  It is applied only after the simultaneous
open-string reduction and collinear expansion have been performed.
It retains the value of the effective kernel on the collinear surface,
\(c_{\rho,0}\), while omitting its off-collinear variation multiplying
the already known leading pole.

\subsection{Hierarchy at strict collinearity}

Using the strict prescription \req{eq:strict-projection-main}, the results
in section \ref{Results} verify the following structure through \(n=9\) and motivate
its all-multiplicity extension.
At the symmetric point \(x=\tfrac12\), the \((n-3)!\) strict coefficients
\(C_\rho^{(0)}\) split into \(n-3\) complementary sector families:
\begin{equation}
 \label{eq:set1}
 \begin{aligned}
 \alpha'^{\,2k-1}
 \sum_{\rho\in S_{n-3}}
 \Re c_\rho^{(2k-1)}\,C_\rho^{(0)}
 &=
 \frac{2g^2}{\kappa}\,
 A_{\rm EYM}^{(2k-2)}(1,\ldots,n-2;P),
 &&
 k=1,\ldots,
 \left\lceil\frac{n-3}{2}\right\rceil,
 \\[1mm]
 \alpha'^{\,2k}
 \sum_{\rho\in S_{n-3}}
 \Im c_\rho^{(2k)}\,C_\rho^{(0)}
 &=0,
 &&
 k=1,\ldots,
 \left\lfloor\frac{n-3}{2}\right\rfloor .
 \end{aligned}
\end{equation}
Here \(A_{\rm EYM}^{(0)}\equiv A_{\rm EYM}\), while
\(A_{\rm EYM}^{(2j)}\), \(j\geq1\), denotes the corresponding
higher-derivative EYM amplitude defined by the same strict projection.
The kernels \(c_\rho^{(k)}\) are homogeneous polynomials of degree
\(k\) in the dimensionful field-theory invariants
\(s_{ij}^{(0)}\).  Equivalently, the combinations
\(\alpha'^k c_\rho^{(k)}\) are homogeneous polynomials of degree
\(k\) in the dimensionless variables
\(s_{ij}=\alpha's_{ij}^{(0)}\) defined in \req{Mandelstam}.

Equation \req{eq:set1} is evaluated on the strict collinear surface.  In
particular, at \(x=\frac12\),
\begin{equation}
 s_{i,n-1}=s_{i,n}
 =\frac{\alpha'}{2}\,s_{iP}^{(0)},
 \qquad
 s_{n-1,n}=0.
 \label{mandelO}
\end{equation}
At the leading field-theory order, the \(1/\epsilon\) contribution to the
complete open--closed combination cancels according to
\req{eq:leading-pole-bcj-cancellation}.  At higher string orders,
\req{eq:set1} does not assert that every pole of the unprojected mixed
disk amplitude vanishes.  Rather, by definition of the strict projection \req{eq:strict-projection-main},
every Yang--Mills entry in \req{eq:set1} denotes only the strict finite
coefficient
\begin{equation}
 A_{\rm YM}(1,\rho,n-1,n)\Big|_{\rm strict,\,\epsilon^0}
 =
 A_{\rm YM}(1,\rho,n-1,n)_{\rm sub}
 =
 C_\rho^{(0)}.
 \label{solvefor}
\end{equation}
Thus no derivative of the off-collinear kernel and no higher coefficient
in its \(\epsilon\)-expansion defines an additional \(\alpha'\)-sector.

Equation  \req{eq:set1} represents $n-3$ sector families of equations. Taking the independent permutations of the gluon labels generates the full Stirling-organized system  involving quadratic, quartic, sextic, etc. BCJ-like relations and
$F^4$, $F^6$, ... corrected EYM amplitudes $A_{\rm EYM}^{(2j)}$.
In total it allows us to express all
$(n-3)!$ subleading collinear subamplitudes \req{solvefor} in terms of a basis of
$\tfrac12(n-3)!$ independent ordinary $A_{\rm EYM}(1,\ldots,n-2;\,P)$ and higher-derivative EYM amplitudes
$A_{\rm EYM}^{(2j)}(1,\ldots,n-2;\,P),\;j\geq1$.

\section{Higher--order  BCJ relations and EYM corrections}\label{Results}

We now apply the momentum-conserving expansion and strict projection
defined in section~\ref{strictprojection} to the explicit open--closed
disk relations. We first discuss the five-, six- and
seven-point cases and then extract the generic pattern.

\subsection{Five--point subleading collinear limits}

In the following we shall consider the $n=5$ case \req{finalf}  describing the scattering of three open and one closed string state.

\paragraph{Order $\alpha'^1$ (real): EYM amplitude.}
The $O(\alpha'^1)$ part of \req{finalf} is real and given\footnote{Alternatively, we may express (\ref{eq:O_alpha1}) in terms of a basis with
non-adjacent legs $4$ and $5$:
\begin{equation}
\begin{aligned}
A(1,2,3;\,q_1,q_2) 
&= -\pi\alpha' \frac{1}{s_{23}s_{45}}\Bigl\{s_{25}[s_{34}s_{13}+s_{24}(s_{13}+s_{23})]\,A_{\rm YM}(1,3,4,2,5) \\
&-s_{35}[s_{34}(s_{12}+s_{23})+s_{24}s_{12}] \,A_{\rm YM}(1,2,4,3,5)\Bigr\} + O(\alpha'^2)\,.
\end{aligned}
\label{altern}
\ee}
by:
\begin{equation}
A(1,2,3;\,q_1,q_2) 
= \pi\; \Bigl[s_{34}\,A_{\rm YM}(1,2,3,4,5) 
- s_{24}\,A_{\rm YM}(1,3,2,4,5)\Bigr] + O(\alpha'^2)\,.
\label{eq:O_alpha1}
\end{equation}
In the leading collinear limit \req{leadingcoll} this expression reproduces
the EYM amplitude \req{eq:EYM}.
After applying BCJ relations the latter gives \cite{Stieberger:2014hba, Stieberger:2015kia_PLB}:
\be\label{eq:subleading_n5} 
A_{\rm EYM}(1,2,3;P)=-\frac{\pi}{2}\; s_{24}\ A_{\rm YM}(1,5,2,4,3)\ .
\ee
With \req{CollExpa} and  $A(1,2,3,4,5)_{\rm sub}=A(1,2,3,4,5)\big|_{\epsilon^0}$ and $A(1,3,2,4,5)_{\rm sub}=A(1,3,2,4,5)\big|_{\epsilon^0}$ for $n=5$ the two collinear expansions \req{ExpaAmp} are:
\begin{align}\label{SYMs}
A_{\rm YM}(1,2,3,4,5)&=\frac{C_{(23)}^{(-1)}}{\epsilon}+\epsilon^0\; A(1,2,3,4,5)_{\rm sub}+ O(\epsilon)\ ,\\
A_{\rm YM}(1,3,2,4,5)&=\frac{C^{(-1)}_{(32)}}{\epsilon}+\epsilon^0\; A(1,3,2,4,5)_{\rm sub}+ O(\epsilon)\ .\label{SYMsa}
\end{align} 
Substituting the latter into (\ref{eq:O_alpha1})  together with
the collinear parametrization \req{MandelO}
\begin{equation}
s_{34} = x\,s_{3P},\qquad
s_{24}= x\,s_{2P},\label{eq:n5_parametrization}
\end{equation}
yields  the EYM amplitude\footnote{We suppress the coupling factor $\tfrac{\kappa}{g^2}$.} \req{eq:EYM}
\begin{equation}
A_{\rm EYM}(1,2,3;P)\big|_{\alpha'^1} 
= \pi\; x\Big[s_{3P}\,A(1,2,3,4,5)_{\rm sub}
-s_{2P}\,A(1,3,2,4,5)_{\rm sub}\Big] \ \epsilon^0\ ,
\label{EYM5}
\end{equation}
which is  independent of $\epsilon$.
Note that in (\ref{EYM5}) the leading $\epsilon^{-1}$ orders of the SYM amplitudes (\ref{SYMs}) and (\ref{SYMsa}) cancel 
subject to the BCJ relation \cite{BCJ}:
\begin{equation}
s_{34}\ C_{(23)}^{(-1)}-s_{24}\ C_{(32)}^{(-1)}\sim s_{34}\ A_{\rm YM}(1,2,3,4)-s_{24}\ A_{\rm YM}(1,3,2,4)=0\ ,
\end{equation}
and the term in the bracket of (\ref{EYM5})  is $O(\epsilon^0)$. On the other hand, in (\ref{eq:O_alpha1}) subject to (\ref{MandelO}) the subleading 
contributions \req{MandelO} of the coefficients $s_{34}$ and $s_{24}$ conspire with the leading 
order of $A_{\rm YM}$ to unwanted contributions of order $\epsilon^0$.
These contributions are the pole-mixing term $\mathcal B$
defined in eq. \req{eq:pole-mixing}. Applying the strict projection \req{eq:strict-projection-main} removes
this known off-collinear contribution and leaves eq. \req{EYM5}, which
depends only on the strict coefficients $C_\rho^{(0)}$.

Hence, following the prescription of section \ref{strictprojection}, we evaluate the EYM
amplitude (\ref{EYM5}) using the hard SYM amplitudes
$A(1,2,3,4,5)_{\rm sub}$ and $A(1,3,2,4,5)_{\rm sub}$, with the
universal leading collinear factorization terms removed. This
prescription will also be applied at higher multiplicity $n$.

\paragraph{Order $\alpha'^2$ (imaginary): stringy correction.}
The $O(\alpha'^2)$ part of \req{finalf} is purely imaginary and given by:
\begin{align}
A(1,2,3;\,q_1,q_2)\ \big|_{\alpha'^2} &= i\pi^2\;\Bigl[\ 
s_{34}(s_{23}+s_{34}-2s_{51})\,A_{\rm YM}(1,2,3,4,5)
\nonumber\\
&\qquad\qquad\quad
+(s_{23}+s_{34}-s_{51})^2\,A_{\rm YM}(1,3,2,4,5)\ 
\Bigr] \ .
\label{eq:O_alpha2}
\end{align}
This defines a stringy correction to the EYM amplitude \req{eq:EYM}. Note that it does not receive higher order corrections  $F^4$ from expanding the open superstring amplitudes \req{EXPA}
in \req{finalf}. Similar to the observation above in the collinear limit it
also develops mixings between leading orders from $A_{\rm YM}$ and subleading orders from the kinematic invariants (\ref{MandelO}). 
With the  collinear parametrization  (\ref{MandelO})
\begin{align}
s_{51} &= -(1-x)(s_{2P}+s_{3P}),\\
s_{34}(s_{23}+s_{34}-2s_{51})&=x\,s_{3P}\Big[(1-2x)s_{2P}+(1-x)s_{3P}\Big],\\
(s_{23}+s_{34}-s_{51})^2 &= x^2 s_{2P}^2.
\label{eq:n5_square_coeff_reduced}
\end{align}
we want to subtract the latter in the following way:
\begin{align}
\hat{A}_{\rm EYM}(1,2,3;\,P) &:=i\pi^2\;\Bigl\{
-s_{34}(s_{24}+s_{51})\,\left[A_{\rm YM}(1,2,3,4,5)-\frac{C_{(23)}^{(-1)}}{\epsilon}\right]
\nonumber\\
&\qquad\qquad\quad
+s_{24}^2\,\left[A_{\rm YM}(1,3,2,4,5)-\frac{C_{(32)}^{(-1)}}{\epsilon}\right]
\Bigr\}\nonumber\\
&=i\pi^2\;\Bigl\{
-xs_{3P}[xs_{2P}+(1-x)s_{1P}]\;\epsilon^0 A(1,2,3,4,5)_{\rm sub}\nonumber\\
&+x^2s_{2P}^2\;\epsilon^0 A(1,3,2,4,5)_{\rm sub}
\Bigr\}
\label{eq:AtildeEYM_n5}
\end{align}

Solving (\ref{EYM5}) and (\ref{eq:AtildeEYM_n5}) for $A(1,2,3,4,5)_{\rm sub},A(1,3,2,4,5)_{\rm sub}$ gives:
\begin{align}
A(1,2,3,4,5)_{\rm sub}&=-\frac{1}{s_{3P}s_{1P}}\ \frac{1}{x(1-x)}\ \left\{xs_{2P}A_{\rm EYM}+\hat A_{\rm EYM}\right\}\ ,\\
A(1,3,2,4,5)_{\rm sub}&=-\frac{1}{s_{2P}s_{1P}}\ \frac{1}{x(1-x)}\ \left\{[xs_{2P}+(1-x)s_{1P}]A_{\rm EYM}+\hat A_{\rm EYM}\right\}\ .
\end{align}

To summarize, although the exact open--closed string relations \req{finalf} are perfectly regular, their strict $\epsilon^0$ projection is subtle. The reason is that subleading terms (\ref{MandelO}) in the kinematic coefficients combine with the universal leading collinear poles of the YM amplitudes and generate additional finite contributions. Therefore the strict subleading system is naturally formulated in terms of hard YM amplitudes with the universal leading collinear factorization terms removed. At order $\alpha'^2$, the mixed open--closed amplitude itself furthermore develops a string-corrected leading factorization residue, whose subtraction defines the finite amplitude $\hat A_{\rm EYM}$.

At the symmetric point $x=1/2$ we find in agreement with \req{NYC}
\be
\hat A_{\rm EYM}=0\\ .
\ee
As a consequence we have:
\begin{align}
A(1,2,3,4,5)_{\rm sub}&=-\frac{2}{s_{3P}s_{1P}}\   s_{2P}\ A_{\rm EYM}(1,2,3;P)\ ,\label{nice5a}\\
A(1,3,2,4,5)_{\rm sub}&=\frac{2}{s_{2P}s_{1P}}\ s_{3P}\ A_{\rm EYM}(1,2,3;P)\ ,\label{nice5b}
\end{align}
i.e. at $n=5$ all subleading collinear limits can be expressed by the  EYM amplitude $A_{\rm EYM}(1,2,3;P)$ given in \req{eq:EYM}. Of course, eq. \req{nice5b} simply follows from \req{nice5a}  by permuting the labels $2$ and $3$ and applying $A_{\rm EYM}(1,3,2;P)=-A_{\rm EYM}(1,2,3;P)$.


\subsection{Six--point subleading collinear limits}
Let us now discuss the six subleading corrections $A(1,\rho(2,3,4),5,6),\ \rho\in S_3$ to the six--point SYM amplitude. They appear in the EYM amplitude \req{eq:EYM6} involving one graviton and four gluons:
\begin{align}
A_{\rm EYM}(1,2,3,4;P)&=\pi\ x\ \big\{\ s_{4P}\;A(1,2,3,4,5,6)_{\rm sub}+s_{2P}\;A(1,4,3,2,5,6)_{\rm sub}\nonumber\\
&-s_{3P}\;[A(1,2,4,3,5,6)_{\rm sub}+A(1,4,2,3,5,6)_{\rm sub}]\ \big\}\ .\label{EYM6}
\end{align}
Note that in the amplitude  \req{EYM6} the leading collinear contributions of the six--point gluon subamplitude cancel as a matter of BCJ relations and only their subleading collinear parts $A_{\rm sub}$ contribute.
Furthermore, at $\ap^2$ from \req{FINAL} we have the imaginary amplitude:
\begin{align}
\widehat{A}_{\rm EYM}(1,2,3,4;P)&=i\pi^2 x\ \Big\{-s_{4P}\;[\;x(s_{2P}+s_{3P})+(1-x)s_{1P}\;]\;A(1,2,3,4,5,6)_{\rm sub}\nonumber\\
&-xs_{2P}^2\;A(1,4,3,2,5,6)_{\rm sub}\nonumber\\
&-[\;s_{34}-x(s_{2P}-s_{4P})-(1-x)s_{1P}\;]\;s_{3P}\; A(1,2,4,3,5,6)_{\rm sub}\nonumber\\
&-[\;s_{34}+s_{24}-x(s_{2P}-s_{4P})-(1-x)s_{1P}\;]\;s_{3P}\; A(1,4,2,3,5,6)_{\rm sub}\Big\} .\label{hatEYM6}
\end{align}
Following the prescription explained in section~\ref{strictprojection}, the subleading collinear corrections are taken into account in the gluon subamplitudes appearing in \req{hatEYM6}.

In order to set up a system of equations  we consider the permutation of gluon
labels $2\leftrightarrow3$  in \req{EYM6}:
\begin{align}
A_{\rm EYM}(1,3,2,4;P)&=\pi\ x\ \Big\{\ s_{4P}\;A(1,3,2,4,5,6)_{\rm sub}+s_{3P}\;A(1,4,2,3,5,6)_{\rm sub}\nonumber\\
&-s_{2P}\;[A(1,3,4,2,5,6)_{\rm sub}+A(1,4,3,2,5,6)_{\rm sub}]\ \Big\}\ .\label{EYM6a}
\end{align}
We have the reflection identity describing KK relations for the gluon labels \cite{Stieberger:2016lng}:
\be
A_{\rm EYM}(1,2,3,4;P)+A_{\rm EYM}(1,3,2,4;P)+A_{\rm EYM}(1,3,4,2;P)=0\ .
\ee
As a consequence, no more independent equations can be gained from permuting $3\leftrightarrow4$ in \req{EYM6}. 
There are only two independent EYM subamplitudes.

Again for $x=1/2$ we have \req{NYC}
\be\label{hatNYC}
\hat A_{\rm EYM}(1,2,3,4;P)=0\ ,
\ee
and for all its permutations in labels $2,3$ and $4$. These relations give 
another set of three independent equations for the six subleading corrections $A(1,\sigma(2,3,4),5,6)_{\rm sub},\ \sigma\in S_3$. We have verified that the
NMHV split--amplitudes \req{NMHVa} satisfy (\ref{hatNYC}). Together with \req{EYM6a} and its permutations we find that five of the six independent subleading collinear limits can be expressed in terms of the three--dimensional basis
\be\label{Set6a}
A_{\rm EYM}(1,2,3,4;P)\ \ \ ,\ \ \ A_{\rm EYM}(1,3,2,4;P)
\ee
and 
\be\label{Set6b}
A(1,2,3,4,5,6)_{\rm sub}\ .
\ee
Hence all six subleading collinear limits of the subamplitudes $A(1,\sigma(2,3,4),5,6)_{\rm sub}$ can be expressed as linear combinations of the three--dimensional basis elements \req{Set6a} and \req{Set6b} with rational coefficients in the kinematic invariants. Thus, in contrast to the $n=5$ case the set of two independent EYM amplitudes \req{Set6a} is not enough to specify all subleading collinear subamplitudes.

Moreover, with this three parameter solution 
all higher order imaginary parts
\be
\lf. \Im A_{\rm EYM}(1,2,3,4;P)\ri|_{\ap^{k}}=0\ \ \ ,\ \ \ k\geq2\ ,
\ee
are automatically solved. We have checked this up to $\ap^{10}$.
To eliminate \req{Set6b} by a gravitational amplitude we consider the 
$\ap^2$ correction to the EYM amplitude
\begin{align}
A^{F^4}_{\rm EYM}(1,2,3,4;P)&=\lf. \Re 
A_{\rm EYM}(1,2,3,4;P)\ri|_{\ap^{3}}\nonumber\\
&=\zeta_2\ \Big\{\ (s_{34}+s_{45})(s_{12}+s_{51})\ A_{\rm EYM}(1,2,3,4;P)\nonumber\\
&+s_{13}s_{24}\ A_{\rm EYM}(1,3,2,4;P)\nonumber\\ 
&-\frac\pi2(s_{12}+s_{34})s_{45}s_{51}\ A(1,2,3,4,5,6)_{\rm sub}\ \Big\}\ .\label{CorrEYM6}
\end{align}
This contribution is due to higher order gauge interactions $\zeta_2\;F^4$ contributing to the EYM amplitude. The amplitude \req{CorrEYM6} allows us to express the subleading correction as
\begin{align}
    A(1,2,3,4,5,6)_{\rm sub}&=\frac2\pi[s_{123}s_{234}(s_{12}+s_{34})]^{-1}\ 
    \Big\{
(s_{14}+s_{24})(s_{13}+s_{14})\ A_{\rm EYM}(1,2,3,4;P)\nonumber\\
&+s_{13}s_{24}\ A_{\rm EYM}(1,3,2,4;P)-\zeta_2^{-1}\; A^{F^4}_{\rm EYM}(1,2,3,4;P)\Big\}\ .\label{solveEYM6}
\end{align}
Of course, the other five subleading collinear limits of the subamplitudes  $A(1,\sigma(2,3,4),5,6)_{\rm sub}$ simply follow from \req{solveEYM6} by respective permutations of gluon labels.

To summarize, for $n=6$ we can express all subleading collinear corrections
in terms of the three gravitational amplitudes \req{Set6a} and \req{CorrEYM6}:
\be
A_{\rm EYM}(1,2,3,4;P)\ \ \ ,\ \ \ A_{\rm EYM}(1,3,2,4;P)\ \ \ ,\ \ \ 
A^{F^4}_{\rm EYM}(1,2,3,4;P)\ .
\ee
All expressions in the six-point reconstruction depend only on
Mandelstam invariants constructed from the hard momenta and the parent
momentum \(P\); the reference spinor \(r\) has disappeared completely.  In particular, the cancellation
occurs for every reconstructed coefficient
\(A(1,\rho(2,3,4),5,6)_{\rm sub}\) separately and is not merely a
cancellation within the EYM combinations, cf. also (\ref{NMHVa}).

\subsection{Seven--point subleading collinear limits}

At seven points we work directly at the symmetric point $x=1/2$ and in
strict collinear kinematics.  There are $4!=24$ subleading collinear
amplitudes
\be
A(1,\rho(2,3,4,5),6,7)_{\rm sub}\ ,\qquad \rho\in S_4\ .
\ee
For illustration, one of the six ordinary EYM equations contained in the
seven--point system is given by
\begin{align}
A_{\rm EYM}(1,2,3,4,5;P)=\frac{\pi}{2}\Big\{
&(s_{12}+s_{13}+s_{14}+s_{23}+s_{24}+s_{34})
 A(1,2,3,4,5,6,7)_{\rm sub}\nonumber\\
&+(s_{12}+s_{23}+s_{24}+s_{25})
 A(1,5,4,3,2,6,7)_{\rm sub}\nonumber\\
&-(s_{13}+s_{23}+s_{34}+s_{35})\nonumber\\[-1mm]
&\quad\times\big[A(1,2,5,4,3,6,7)_{\rm sub}
     +A(1,5,2,4,3,6,7)_{\rm sub}\nonumber\\[-1mm]
&\hspace{23mm}+A(1,5,4,2,3,6,7)_{\rm sub}\big]\nonumber\\
&+(s_{14}+s_{24}+s_{34}+s_{45})\nonumber\\[-1mm]
&\quad\times\big[A(1,2,3,5,4,6,7)_{\rm sub}
     +A(1,2,5,3,4,6,7)_{\rm sub}\nonumber\\[-1mm]
&\hspace{23mm}+A(1,5,2,3,4,6,7)_{\rm sub}\big]\Big\}\ ,
\label{EYM7a}
\end{align}
The remaining five equations follow from the other orderings of the five
gluon labels.  The numbers of independent constraints obtained from the $\ap$ expansion at orders 1 through 4 are 
\be
6+11+6+1=24\ .
\ee
Here the first and third terms are supplied by ordinary and $F^4$-corrected
EYM amplitudes, respectively, while the second and fourth terms are the
quadratic and quartic homogeneous constraints.  In particular, the even
constraints have total rank twelve, and the remaining twelve quantities can
be chosen to be gravitational amplitudes.

A convenient choice for the six ordinary EYM amplitudes is
\begin{align}
\{&A_{\rm EYM}(1,2,3,4,5;P),\ A_{\rm EYM}(1,2,4,3,5;P),
   \ A_{\rm EYM}(1,3,2,4,5;P),\nonumber\\
 &A_{\rm EYM}(1,3,4,2,5;P),\ A_{\rm EYM}(1,4,2,3,5;P),
   \ A_{\rm EYM}(1,4,3,2,5;P)\}\ ,
\end{align}
while the six independent $F^4$-corrected amplitudes may be chosen as
\begin{align}
\{&A^{F^4}_{\rm EYM}(1,2,3,4,5;P),\ A^{F^4}_{\rm EYM}(1,2,3,5,4;P),
   \ A^{F^4}_{\rm EYM}(1,2,4,3,5;P),\nonumber\\
 &A^{F^4}_{\rm EYM}(1,2,4,5,3;P),\ A^{F^4}_{\rm EYM}(1,2,5,3,4;P),
   \ A^{F^4}_{\rm EYM}(1,2,5,4,3;P)\}\ .
\end{align}
Solving the complete system expresses every one of the 24 orderings in terms
 of  these twelve amplitudes. 

As a compact representative of the solution, the canonical ordering is
\be\label{SevenPointCanonical}
A(1,2,3,4,5,6,7)_{\rm sub}
=\frac{2}{\pi}\frac{\mathcal N_{12345}}{s_{1234} s_{2345}(s_{123}+s_{45})(s_{12}+s_{345}) }\ ,
\ee
where
\begin{align}
\mathcal N_{12345}={}& \; (4\zeta_2)^{-1} \big\{ A^{F^4}_{\rm EYM}(1,2,5,4,3;P)
 (2s_{13}-s_{14}+s_{24}-2s_{25}-s_{35})\nonumber\\
&- \; A^{F^4}_{\rm EYM}(1,2,5,3,4;P)\;
 (2s_{13}-s_{14}+s_{24}+2s_{25}-s_{35})\nonumber\\
&+ \;A^{F^4}_{\rm EYM}(1,2,3,5,4;P)\;
 (2s_{13}+3s_{14}+s_{24}+2s_{25}-s_{35})\nonumber\\
&+ \;A^{F^4}_{\rm EYM}(1,2,4,5,3;P)\;
 (2s_{13}+s_{14}-s_{24}+2s_{25}+s_{35})\nonumber\\
&- \;A^{F^4}_{\rm EYM}(1,2,4,3,5;P)\;
 (2s_{13}+s_{14}+3s_{24}+2s_{25}+s_{35})\nonumber\\
&- \;A^{F^4}_{\rm EYM}(1,2,3,4,5;P)\nonumber\\
&\quad\times(4s_{12}+2s_{13}+3s_{14}+4s_{23}+s_{24}
  +2s_{25}+4s_{34}+3s_{35}+4s_{45})\big\}\nonumber\\
&-A_{\rm EYM}(1,4,3,2,5;P)\;s_{14}s_{25}
 (s_{12}+s_{23}+s_{34}+s_{45})\nonumber\\
&+A_{\rm EYM}(1,3,4,2,5;P)\;s_{13}s_{25}
 (s_{12}+s_{13}+s_{14}+s_{23}+s_{34}+s_{45})\nonumber\\
&+A_{\rm EYM}(1,3,2,4,5;P)\;s_{13}(s_{24}+s_{25})\nonumber\\
&\quad\times
 (s_{12}+s_{13}+s_{14}+s_{23}+s_{24}+s_{34}+s_{45})\nonumber\\
&A_{\rm EYM}(1,4,2,3,5;P)\;s_{14}s_{35}
 (s_{12}+s_{23}+s_{25}+s_{34}+s_{35}+s_{45})\nonumber\\
&+A_{\rm EYM}(1,2,4,3,5;P)\;(s_{14}+s_{24})s_{35}\nonumber\\
&\quad\times
 (s_{12}+s_{23}+s_{24}+s_{25}+s_{34}+s_{35}+s_{45})\nonumber\\
&A_{\rm EYM}(1,2,3,4,5;P)\;
 (s_{12}+s_{13}+s_{14}+s_{23}+s_{24}+s_{34}+s_{45})\nonumber\\
&\quad\times
 (s_{12}+s_{13}+s_{23}+s_{34}+s_{35}+s_{45})\nonumber\\
&\quad\times
 (s_{12}+s_{23}+s_{24}+s_{25}+s_{34}+s_{35}+s_{45})\ .
\end{align}
The other 23 orderings have the same rational structure and involve only the
twelve amplitudes displayed above.


\subsection{Generic $n$}

The explicit low-point examples above constitute the first members of a
general hierarchy.  We have constructed the corresponding open--closed
coefficient blocks and computed their ranks at generic kinematics for all
multiplicities $5\leq n\leq 9$.  In every case, the $(n-3)!$ strict
subleading collinear data decompose into $n-3$ mutually independent
sectors whose ranks are given by the unsigned Stirling numbers:
\be
c(n-3,k)=\left[\begin{matrix} n-3 \\ k \end{matrix}\right]\ ,
\qquad
k=1,\ldots,n-3\ .
\ee
The results verified through $n=9$ are summarized in table~1, together
with the conjectural continuation to $n=10$.

The even sectors add up to one half of the total number of YM data,
\be
\sum_{j\geq 1}
\left[{n-3\atop 2j}\right]
=
\frac12 (n-3)!,
\ee
and are identified with higher-order BCJ-like relations. The odd sectors add
up to the complementary half,
\be
\sum_{j\geq 0}
\left[{n-3\atop 2j+1}\right]
=
\frac12 (n-3)!,
\ee
and provide the gravitational basis consisting of ordinary EYM amplitudes and
higher-derivative EYM corrections.

\begin{table}[H]
\renewcommand{\arraystretch}{1.25}
\hskip-1.5cm
\resizebox{1.2 \textwidth}{!}{
\begin{tabular}{|c|c|c|c|c|c|c|c|c|c|}
\hline
$n$ 
& $(n-3)!$
& $\tfrac12(n-3)!$
& $\left[{n-3\atop 1}\right]=(n-4)!$
& $\left[{n-3\atop 2}\right]$
& $\left[{n-3\atop 3}\right]$
& $\left[{n-3\atop 4}\right]$
& $\left[{n-3\atop 5}\right]$
& $\left[{n-3\atop 6}\right]$
& $\left[{n-3\atop 7}\right]$
\\
\hline
&
&
&
EYM
&
quadratic
&
$F^4$--EYM
&
quartic
&
$F^6$--EYM
&
sextic
&
$F^8$--EYM
\\
&
&
&
\ 
&
BCJ-like
&
\ 
&
BCJ-like
&
\ 
&
BCJ-like
&
\ 
\\
\hline
&
&
&
$\ap\;\pi$ &
$\ap^2\;i\pi^2$ &
$\ap^3\;\pi\zeta_2$&
$\ap^4\;i\pi^4$ &
$\ap^5\;\pi\zeta_2^2$&
$\ap^6\;i\pi^6$&
$\ap^7\;\pi\zeta_2^3$\\
\hline
$5$ & $2$ & $1$  & $1$   & $1$   & --   & --   & --  & -- & --
\\
$6$ & $6$    & $3$ & $2$   & $3$   & $1$   & --   & --  & -- & --
\\
$7$ & $24$ & $12$ & $6$   & $11$  & $6$   & $1$  & --  & -- & --
\\
$8$ & $120$ & $60$& $24$  & $50$  & $35$  & $10$ & $1$ & -- & --
\\
$9$ & $720$ &$360$ & $120$ & $274$ & $225$ & $85$ & $15$ & $1$ & --
\\
$10$& $5040$ &$2520$& $720$&$1764$&$1624$&$735$&$175$&$21$&$1$\\
\hline
\end{tabular}
}
\caption{
Decomposition of the $(n-3)!$ strict subleading collinear data according to
unsigned Stirling numbers of the first kind,
$c(n-3,k)=\left[{n-3\atop k}\right]$.
For $n=10$ the pattern needs still to be verified.
}
\label{tab:stirling-counting}
\end{table}

Albeit the columns in table \ref{tab:stirling-counting} describe consecutive orders in the $\ap$-expansion, they label independent Stirling sectors of the projected subleading-collinear data. A given order in $\ap$ may contain several transcendental structures, and these structures may either generate a new sector or project to the span of sectors already present at lower order.
The refined pattern suggests a sharper interpretation. The odd Stirling sectors are represented by real EYM-type data,
$$A_{\rm EYM}^{(0)},\quad A_{\rm EYM}^{(2)},\quad A_{\rm EYM}^{(4)},\ldots,$$
whose leading effective-field-theory contents are of Born–Infeld type, schematically as
$$F^2,\quad F^4,\quad F^6,\ldots\ .$$
The couplings $F^{2k}$ comprise also independent terms of the sort $D^{2l}F^{2k-l}$.
The even Stirling sectors are represented by homogeneous BCJ-like constraints generated by the imaginary open–closed monodromy projections \req{NYC}. Thus, table \ref{tab:stirling-counting} counts independent projected data, not independent local operators at consecutive orders in $\ap$.
This also clarifies the role of odd zeta values. The matrices $M_{2r+1}$ in the open string expansion \req{EXPA} are present in the full open-string period matrix, but in the present open–closed projection \req{FINAL} they do not appear to generate new independent sectors of the strict subleading-collinear quotient. Instead, the new gravitational data are carried by the even-zeta/Born–Infeld part $P_{2k}$ of the expansion, whereas the odd-zeta matrices $M_{2k+1}$ act within the span already generated by the lower sectors or are removed by the monodromy reality constraints at $q_1=q_2$, i.e.
\be\label{half}
x=\fc12\ ,
\ee
according to \req{leadingcoll}.

The Stirling-number pattern should not be attributed to the full open-string amplitude alone. Rather, it appears after applying the open–closed monodromy projection in \req{FINAL}, whose sine factors and complex phases select specific real and imaginary combinations of open-string partial amplitudes. At the symmetric point $q_1=q_2$, the imaginary projections vanish \req{NYC}, but leave non--trivial homogeneous constraints. Thus, the particular \(\left[{n-3\atop k}\right]\) decomposition found here is a property of the BCJ-reduced open–closed projection. On the other hand,  the unrestricted open-string amplitudes exhibit the related but different \(\left[{N-1\atop k}\right]\) descent-algebra organization of Ref. \cite{Mafra:2021wok}. Hence, our Stirling decomposition seems to be a property of the projected open–closed monodromy kernels \req{kernel}, not of the unrestricted open-string period matrix $F$.

\section{Higher--order BCJ relations and gravitational basis}

The examples of the preceding section \ref{Results} show that ordinary EYM amplitudes do
not contain enough information to determine all strict subleading collinear
YM coefficients.  The missing information is supplied by the higher orders
of the open--closed disk relation \req{FINAL}.  At every polynomial degree
$k$ they generate a new block of equations with kinematic coefficients of
degree $k$.  Even-degree blocks give homogeneous BCJ-like relations, whereas
odd-degree blocks relate the YM coefficients to ordinary or
higher-derivative EYM amplitudes.

\subsection{The complete linear system}

Collect the $(n-3)!$ pole-subtracted strict coefficients defined in
\req{solvefor} into the column vector
\begin{equation}
 {\boldsymbol C}_n
 :=
 \bigl(C_\rho^{(0)}\bigr)_{\rho\in S_{n-3}},
 \qquad
 C_\rho^{(0)}
 :=
 A_{\rm YM}(1,\rho,n-1,n)_{\rm sub.}\, .
 \label{strict-data-vector}
\end{equation}
All statements below are understood at generic kinematics \req{Mandelstam} subject to \req{leadingcoll} and \req{half}.

For each degree $k=1,\ldots,n-3$, we apply the relevant
relabellings to $\Re c_\rho^{(k)}$ for odd $k$ and to
$\Im c_\rho^{(k)}$ for even $k$, as prescribed by \req{eq:set1}.
We proceed successively in increasing degree. At degree $k$, we retain
a maximal set of coefficient rows that is linearly independent not
only within that degree, but also of all rows already retained at
degrees $1,\ldots,k-1$. We denote the resulting matrix of genuinely
new rows by $W_n^{(k)}$; its columns are labelled by
$\rho\in S_{n-3}$.  After stripping off the common
power $\alpha'^k$, the corresponding relations \req{eq:set1} can be written as
\begin{equation}
 \begin{aligned}
  W_n^{(2j)}{\boldsymbol C}_n&=0,
     &&1\leq 2j\leq n-3\ ,\\[1mm]
  W_n^{(2j+1)}{\boldsymbol C}_n
     &={\boldsymbol A}_{{\rm EYM},n}^{(2j)}\ ,
     &&1\leq 2j+1\leq n-3 .
 \end{aligned}
 \label{block-system}
\end{equation}
Here ${\boldsymbol A}_{{\rm EYM},n}^{(2j)}$ collects the EYM amplitudes
associated with the same selected rows, including all order-dependent coupling and transcendental normalization factors, such as powers of $\pi$ and zeta values.  The case $j=0$ gives ordinary EYM, while $j\geq1$ gives the $F^4$-, $F^6$- and
subsequent higher-derivative corrections.  Although the entries of $W_n^{(k)}$ are degree-$k$ polynomials in Mandelstam invariants \req{Mandelstam}, the
system \req{block-system} is linear in the unknown coefficients $C_\rho^{(0)}$.
Our generic-kinematics rank computations through $n=9$ give
\begin{equation}
 \operatorname{rk}W_n^{(k)}
 =
 \genfrac{[}{]}{0pt}{}{n-3}{k},
 \qquad
 \operatorname{rk}
 \begin{pmatrix}
  W_n^{(1)}\\
  W_n^{(2)}\\
  \vdots\\
  W_n^{(n-3)}
 \end{pmatrix}
 =(n-3)! .
 \label{stirling-dual-decomposition}
\end{equation}
By construction, no row direction already present in the
lower-degree blocks is counted again in $W_n^{(k)}$. The cumulative
row spaces obtained by successively stacking
$W_n^{(1)},W_n^{(2)},\ldots,$
$W_n^{(k)}$ therefore form a filtration,
and $W_n^{(k)}$ represents the genuinely new block appearing at
degree $k$. 
Thus the blocks are mutually independent and together give a complete
system for the $(n-3)!$ strict YM coefficients \req{solvefor}.  We conjecture that
\req{stirling-dual-decomposition} continues to hold at arbitrary
multiplicity $n$.  This is a rank decomposition of the coefficient matrices $W_n^{(k)}$ 
after reduction to a BCJ basis.  It is not a decomposition of individual
YM amplitudes by transcendental weight.  In particular, additional
transcendental contributions at the same or at higher orders in $\alpha'$
need not generate a new block when their coefficient rows lie in the span
of rows already present. This applies to all MZV basis elements that do not contain $\zeta_2$ factors.

To display the physical content, we stack the even and odd matrices
separately:
\begin{equation}
 R_n:=
 \begin{pmatrix}
  W_n^{(2)}\\
  W_n^{(4)}\\
  \vdots
 \end{pmatrix},
 \qquad
 G_n:=
 \begin{pmatrix}
  W_n^{(1)}\\
  W_n^{(3)}\\
  \vdots
 \end{pmatrix},
 \label{even-odd-matrices}
\end{equation}
where only blocks with $k\leq n-3$ are included.  In this notation the full
system \req{block-system} is simply
\begin{equation}
 R_n{\boldsymbol C}_n=0,
 \qquad
 G_n{\boldsymbol C}_n={\boldsymbol A}^{\,\rm grav}_n .
 \label{even-odd-system}
\end{equation}
The vector ${\boldsymbol A}^{\,\rm grav}_n$ contains the independent
ordinary and higher-derivative EYM amplitudes, with the normalization in
\req{block-system} understood.
The parity identities for unsigned Stirling numbers, together with
\req{stirling-dual-decomposition}, imply
\begin{equation}
 \operatorname{rk}R_n
 =
 \operatorname{rk}G_n
 =
 \frac12\,(n-3)!,
 \qquad
 \operatorname{rk}
 \begin{pmatrix}
  R_n\\[1mm]
  G_n
 \end{pmatrix}
 =(n-3)! .
 \label{complete-rank}
\end{equation}
Since the matrices $W_n^{(k)}$ were defined by retaining independent
row bases, the matrices $R_n$ and $G_n$ contain no redundant rows.
Consequently,
\[
 R_n,\ G_n
 \in \mathbb{R}^{\frac{1}{2}(n-3)!\,\times\,(n-3)!}\ ,
\]
whereas their vertical concatenation is a square matrix,
\[
 \begin{pmatrix}
 R_n\\[1mm]
 G_n
 \end{pmatrix}
 \in
 \mathbb{R}^{(n-3)!\times(n-3)!}.
\]
Thus $R_n$ and $G_n$ are individually rectangular, while the
stacked matrix is square and invertible at generic kinematics.
The homogeneous equations therefore reduce the $(n-3)!$ YM coefficients to
$\tfrac12(n-3)!$ independent combinations:
\begin{equation}
 \dim\ker R_n=\frac12\,(n-3)! .
 \label{even-kernel}
\end{equation}
The second equation in \req{even-odd-system} determines these remaining
combinations {\it uniquely}.  Equivalently, the action of $G_n$ on
$\ker R_n$ is invertible at generic kinematics.  This is the precise sense
in which the EYM-type amplitudes $A_{\rm EYM}^{(2j)}$ provide coordinates on the solution
space selected by the even BCJ-like constraints.

\paragraph{Constructive reconstruction.}
Because the matrices in \req{even-odd-matrices} contain independent row
bases, the stacked matrix in \req{complete-rank} is square.  The strict
subleading YM coefficients \req{strict-data-vector} are therefore reconstructed directly as:
\begin{equation}
 {\boldsymbol C}_n
 =
 \begin{pmatrix}R_n\\[1mm]G_n\end{pmatrix}^{-1}
 \begin{pmatrix}0\\[1mm]{\boldsymbol A}^{\,\rm grav}_n\end{pmatrix}.
\label{reconstruction-formula}
\end{equation}
Hence the coefficients of the gravitational expansion are rational
functions of Mandelstam invariants.  They can be obtained directly from
the inverse of the complete coefficient matrix, or from its minors.
Equation \req{reconstruction-formula} has been verified through $n=9$ and
gives an all-multiplicity reconstruction prescription conditional on the
full-rank conjecture.

\paragraph{Low-point realization.}
The matrices \(R_n\) and \(G_n\) depend on the choice and normalization
of the independent rows. We give convenient representatives below.
Kinematics-independent non-singular transformations within the rows of
\(R_n\) or \(G_n\) lead to equivalent reconstruction matrices and do
not affect their generic rank. The determinant formulas below refer to
the fixed homogeneous polynomial normalization inherited from the
open--closed monodromy kernels. Kinematics-dependent row rescalings are
not included, since they would modify the factorization of the
determinant.

At five points, let
\[
 {\boldsymbol C}_5=
 \begin{pmatrix}
 C_{23}^{(0)}\\
 C_{32}^{(0)}
 \end{pmatrix}.
\]
After removing the common normalization factors already absorbed into
\({\boldsymbol A}^{\rm grav}_5\), eqs.~\req{EYM5} and
\req{eq:AtildeEYM_n5}, evaluated at \(x=\frac12\), give
\[
 R_5=
 \begin{pmatrix}
 s_{3P}^{\,2} & s_{2P}^{\,2}
 \end{pmatrix},
 \qquad
 G_5=
 \begin{pmatrix}
 s_{3P} & -s_{2P}
 \end{pmatrix}.
\]
Consequently,
\[
 \begin{pmatrix}
 R_5\\[1mm]
 G_5
 \end{pmatrix}
 {\boldsymbol C}_5
 =
 \begin{pmatrix}
 0\\
 {\cal E}_5
 \end{pmatrix},
\]
where \({\cal E}_5\) denotes the correspondingly normalized EYM
amplitude. The determinant is
\begin{equation}
 {\cal D}_5:=
 \det
 \begin{pmatrix}
 R_5\\[1mm]
 G_5
 \end{pmatrix}
 =
 s_{1P}s_{2P}s_{3P}.
 \label{det-five-point}
\end{equation}
Thus the matrix is invertible at generic kinematics, and its inverse
reproduces eqs.~\req{nice5a} and \req{nice5b}.

At six points, we choose the ordering
\[
 {\boldsymbol C}_6=
 \begin{pmatrix}
 C_{234}^{(0)}&
 C_{243}^{(0)}&
 C_{324}^{(0)}&
 C_{342}^{(0)}&
 C_{423}^{(0)}&
 C_{432}^{(0)}
 \end{pmatrix}^{\!t}
\]\ .
After stripping the common normalization
factors, the two ordinary EYM relations
\req{EYM6} and \req{EYM6a} are represented by the rows
\[
 {\boldsymbol g}_{234}^{(1)}
 =
 \begin{pmatrix}
 s_{4P}&-s_{3P}&0&0&-s_{3P}&s_{2P}
 \end{pmatrix},
\]
and
\[ {\boldsymbol g}_{324}^{(1)}
 =\begin{pmatrix}
 0&0&s_{4P}&-s_{2P}&s_{3P}&-s_{2P}
 \end{pmatrix}.\]
Thus
\[
 {\boldsymbol g}_{234}^{(1)}{\boldsymbol C}_6
 ={\cal E}_{234},
 \qquad
 {\boldsymbol g}_{324}^{(1)}{\boldsymbol C}_6
 ={\cal E}_{324},
\]
where \({\cal E}_{234}\) and \({\cal E}_{324}\) denote the
correspondingly normalized ordinary EYM amplitudes \req{EYM6} and \req{EYM6a}, respectively.
The third gravitational row is obtained directly from the
\(F^4\)-corrected EYM relation \req{CorrEYM6}. Define
\[
 \alpha_6=(s_{14}+s_{24})(s_{13}+s_{14}),
 \qquad
 \beta_6=s_{13}s_{24},
\]
and
\[
 \Delta_6=s_{123}s_{234}(s_{12}+s_{34}),
 \qquad
 {\boldsymbol e}_{234}
 =
 \begin{pmatrix}
 1&0&0&0&0&0
 \end{pmatrix}.
\]
After stripping the common factor \(\zeta_2/2\), and using the
strict \(x=\frac12\) kinematics and momentum conservation,
eq.~\req{CorrEYM6} becomes
\[
 {\cal F}_{234}
 =
 \alpha_6{\cal E}_{234}
 +\beta_6{\cal E}_{324}
 -\Delta_6 C_{234}^{(0)},
\]
with \({\cal F}_{234}\) denoting the correspondingly normalized
\(F^4\)-corrected EYM amplitude \(A_{\rm EYM}^{F^4}(1,2,3,4;P)\). Consequently, its coefficient row is
\[
 {\boldsymbol g}_{234}^{(3)}
 =
 \alpha_6{\boldsymbol g}_{234}^{(1)}
 +\beta_6{\boldsymbol g}_{324}^{(1)}
 -\Delta_6{\boldsymbol e}_{234}.
\]
Explicitly,
\[
 {\boldsymbol g}_{234}^{(3)}
 =
 \begin{pmatrix}
 \alpha_6s_{4P}-\Delta_6&
 -\alpha_6s_{3P}&
 \beta_6s_{4P}&
 -\beta_6s_{2P}&
 (\beta_6-\alpha_6)s_{3P}&
 (\alpha_6-\beta_6)s_{2P}
 \end{pmatrix}.
\]
The gravitational matrix and its right-hand side may therefore be
chosen as
\begin{align*}
 G_6&=
 \begin{pmatrix}
 s_{4P}&-s_{3P}&0&0&-s_{3P}&s_{2P}\\
 0&0&s_{4P}&-s_{2P}&s_{3P}&-s_{2P}\\
 \alpha_6s_{4P}-\Delta_6&
 -\alpha_6s_{3P}&
 \beta_6s_{4P}&
 -\beta_6s_{2P}&(\beta_6-\alpha_6)s_{3P}&(\alpha_6-\beta_6)s_{2P}
 \end{pmatrix},\\
& {\boldsymbol A}^{\rm grav}_6=
 \begin{pmatrix}
 {\cal E}_{234}\\
 {\cal E}_{324}\\
 {\cal F}_{234}
 \end{pmatrix}.
\end{align*}
In this representation, all three entries of
\({\boldsymbol A}^{\rm grav}_6\) are ordinary or higher-derivative
EYM amplitudes; no triangularly shifted gravitational datum is
introduced.
For any permutation \((a,b,c)\) of \((2,3,4)\), we introduce the
local abbreviations
\begin{equation}
 u_{abc}:=2s_{bc}+2s_{cP}+s_{bP},
 \qquad
 v_{abc}:=2s_{bc}+2s_{ac}+2s_{cP}+s_{bP}
         =u_{abc}+2s_{ac}.
 \label{quadratic-six-point-factors}
\end{equation}
In terms of these quantities, and for the column ordering of
\({\boldsymbol C}_6\) displayed above, a convenient independent set of
the three homogeneous quadratic relations is
\begin{equation}
 R_6=\frac12
 \begin{pmatrix}
 s_{4P}^{\,2}
 &
 -s_{3P}u_{234}
 &
 0
 &
 0
 &
 -s_{3P}v_{234}
 &
 -s_{2P}^{\,2}
 \\[1mm]
 0
 &
 0
 &
 s_{4P}^{\,2}
 &
 -s_{2P}u_{324}
 &
 -s_{3P}^{\,2}
 &
 -s_{2P}v_{324}
 \\[1mm]
 -s_{4P}u_{243}
 &
 s_{3P}^{\,2}
 &
 -s_{4P}v_{243}
 &
 -s_{2P}^{\,2}
 &
 0
 &
 0
 \end{pmatrix}.
 \label{explicit-R6}
\end{equation}
The three rows correspond  to the independent
row choice associated with the permutations \((234)\), \((324)\), and
\((243)\) of \req{hatEYM6}.  Other independent choices are related to this matrix by
non-singular row transformations.
Thus \(R_6\) and \(G_6\) are both \(3\times6\) matrices, whereas
\begin{equation}
 {\cal M}_6:=
 \begin{pmatrix}
 R_6\\[1mm]
 G_6
 \end{pmatrix}
 \label{Matrix6}
\end{equation}
is a \(6\times6\) matrix of full rank at generic kinematics.
Using momentum conservation for the five hard momenta,
\[
 s_{123}=s_{4P},
 \qquad
 s_{234}=s_{1P},
 \qquad
 \sum_{a=1}^{4}s_{aP}=0,
\]
we find, for the row and column ordering displayed above,
\begin{equation}
 {\cal D}_6:=\det{\cal M}_6
 =
 \frac14
 \left(\prod_{a=1}^{4}s_{aP}^{\,2}\right)
 (s_{12}+s_{34})(s_{13}+s_{24})(s_{14}+s_{23}) .
 \label{det-complete-six-point}
\end{equation}
The overall constant and sign depend on the normalization and ordering
of the independent rows, whereas the generic non-vanishing and the
kinematic factors in \eqref{det-complete-six-point} are basis
independent within the fixed polynomial normalization.

The inverse of \({\cal M}_6\) is the six-point realization of the
reconstruction formula \req{reconstruction-formula}. Writing out the
inverse is unnecessary, since the component relations required for its
construction are already contained in eqs.~\req{EYM6},
\req{hatEYM6}, \req{EYM6a}, and \req{CorrEYM6}.

\paragraph{Higher-point pattern.}
The matrices above realize the six-point decomposition
\[
 6=2+3+1 ,
\]
with the three blocks \(W_6^{(1)}\), \(W_6^{(2)}\), and
\(W_6^{(3)}\) having ranks \(2\), \(3\), and \(1\), respectively.
At seven points one finds
\[
 24=6+11+6+1 .
\]
The \(11+1=12\) even-sector relations leave a twelve-dimensional
solution space, and the \(6+6=12\) odd-sector amplitudes provide its
gravitational coordinates. At eight points the corresponding
decomposition is
\[
 120=24+50+35+10+1 ,
\]
with \(50+10=60\) even-sector relations and \(24+35+1=60\)
gravitational data. At nine points our computation gives
\[
 720=120+274+225+85+15+1 .
\]
The even blocks contain \(274+85+1=360\) independent BCJ-like
relations. In particular, the one-dimensional \(k=6\) block is a
sextic BCJ-like relation and is included in the combined rank
computation. The odd blocks contain
\(120+225+15=360\) gravitational data: ordinary EYM together with
the \(F^4\)- and \(F^6\)-corrected EYM amplitudes.

\paragraph{Determinant pattern.}
The low-point examples also reveal a simple factorization pattern.
Let
\begin{equation}
 {\cal D}_n:=
 \det
 \begin{pmatrix}
 R_n\\[1mm]
 G_n
 \end{pmatrix}.
 \label{complete-determinant}
\end{equation}
Within the fixed polynomial normalization, kinematics-independent changes of row bases multiply \({\cal D}_n\) only by a non-zero kinematics-independent constant; permutations change only its sign. Hence its generic non-vanishing and its kinematic factorization are basis-independent in this restricted sense. In the homogeneous polynomial normalization inherited from the
open--closed monodromy kernels, \(W_n^{(k)}\) contains
\(\genfrac{[}{]}{0pt}{}{n-3}{k}\) independent rows of kinematic
degree \(k\). Therefore
\begin{equation}
 \deg{\cal D}_n
 =
 \sum_{k=1}^{n-3}
 k\genfrac{[}{]}{0pt}{}{n-3}{k}
 =
 (n-3)!\,H_{n-3}
 =
 \genfrac{[}{]}{0pt}{}{n-2}{2},
 \label{degree-complete-determinant}
\end{equation}
where \(H_{n-3}\) is a harmonic number. This gives
\(\deg{\cal D}_5=3\), \(\deg{\cal D}_6=11\), and
\(\deg{\cal D}_7=50\), in agreement with the explicit results.

For any non-empty subset \(I\subset\{1,\ldots,n-2\}\), we use the
notation
\begin{equation}
 s_I:=\left(\sum_{i\in I}p_i\right)^2,
 \qquad
 I^c:=\{1,\ldots,n-2\}\setminus I .
 \label{multiparticle-invariant}
\end{equation}
An exact evaluation of the complete \(24\times24\) reconstruction
matrix at seven points gives
\begin{equation}
 {\cal D}_7
 \doteq
 \left(\prod_{a=1}^{5}s_{aP}^{\,6}\right)
 \prod_{\substack{I|I^c\ {\rm unordered}\\
 I\cup I^c=\{1,\ldots,5\}\\
 |I|=2}}
 \bigl(s_I+s_{I^c}\bigr)^2 .
 \label{det-seven-point}
\end{equation}
The factorization and all exponents in eq.~\eqref{det-seven-point}
have been verified by an exact evaluation of the complete
\(24\times24\) reconstruction matrix.  Thus eq.~\eqref{det-seven-point}
is not conjectural.  The symbol \(\doteq\) only suppresses a non-zero,
kinematics-independent overall constant that depends on the
normalization and ordering of the selected independent rows. Here and below, \(\doteq\) denotes equality up to a non-zero constant
that depends on the normalization and ordering of the independent
rows. The five factors \(s_{aP}\) correspond to the \(1|4\)
bipartitions of the hard-gluon labels, while the ten factors in the
second product correspond to the \(2|3\) bipartitions. The total
degree is
\[
 5\cdot6+10\cdot2=50,
\]
as required by eq.~\eqref{degree-complete-determinant}.

The results at five, six, and seven points suggest the
all-multiplicity factorization
\begin{equation}
 {\cal D}_n
 \doteq
 \left(
 \prod_{a=1}^{n-2}
 s_{aP}^{\,(n-4)!}
 \right)
 \prod_{\substack{I|I^c\ {\rm unordered}\\
 I\cup I^c=\{1,\ldots,n-2\}\\
 |I|\geq2,\ |I^c|\geq2}}
 \bigl(s_I+s_{I^c}\bigr)^{
 (|I|-1)!(|I^c|-1)!
 } .
 \label{conjectural-Dn}
\end{equation}
The first product displays explicitly the contribution from the
singleton bipartitions. Indeed, for
\(I=\{a\}\), momentum conservation and \(p_a^2=P^2=0\) imply
\[
 s_I+s_{I^c}
 =
 \left(\sum_{i\neq a}p_i\right)^2
 =
 (-P-p_a)^2
 =
 s_{aP},
\]
where the exponent counts the choices of cyclic orderings on the two blocks \(I\) and \(I^c\):
\[
 (|I|-1)!(|I^c|-1)!
 = 0!\,(n-4)!=(n-4)!\ .
\]
Thus eq.~\eqref{conjectural-Dn} is equivalent to a product over all
unordered bipartitions, including the singleton ones, but makes the
individual \(s_{aP}\) factors manifest. For \(n=5\), the second
product in eq.~\eqref{conjectural-Dn} is understood to be empty.
Equation~\eqref{conjectural-Dn} is a conjecture for \(n\geq8\). It is
stated in the polynomial row normalization used above and up to a
non-zero overall constant. Its degree is
\begin{align}
 \deg{\cal D}_n
 &=
 (n-2)(n-4)!
 +\frac12\sum_{r=2}^{n-4}
 \binom{n-2}{r}(r-1)!(n-r-3)!
 \nonumber\\
 &=
 (n-3)!\,H_{n-3}
 =
 \genfrac{[}{]}{0pt}{}{n-2}{2},
 \label{degree-conjectural-Dn}
\end{align}
and hence agrees with the degree fixed independently by the Stirling
decomposition. Our rank computations imply
\({\cal D}_n\not\equiv0\) through \(n=9\); generic non-vanishing at
arbitrary multiplicity remains equivalent to the full-rank conjecture
\eqref{complete-rank}.

\subsection{Relation to known Stirling counting}

The appearance of the $F^4$ sector has an independent precedent.  The
relation between the $\alpha'^2F^4$ correction and one-loop all-plus
field-theory amplitudes was observed and discussed in
\cite{Stieberger:2006bh,Stieberger:2006te}.  The same $F^4$ kinematic
structures occur in one-loop open-superstring and finite equal-helicity
YM amplitudes, where their independent BRST-invariant basis is counted by
unsigned Stirling numbers \cite{Mafra:2012kh}.  This supports the
identification of our $k=3$ block with the $F^4$ Stirling sector.  We are
not aware of an equally direct one-loop field-theory interpretation of
the higher blocks.

Mafra found a related Stirling-number organization in the KK-like
relations of pure-open superstring disk amplitudes
\cite{Mafra:2021wok}.  There the motivic components
$A_{f_2^r f_M}$ are considered modulo relations with rational,
Mandelstam-independent coefficients.  For an $N$-point amplitude the
proposed basis dimensions are
\[
 d_{N,0}=\genfrac{[}{]}{0pt}{}{N-1}{1},
 \qquad
 d_{N,1}=\genfrac{[}{]}{0pt}{}{N-1}{3},
 \qquad
 d_{N,r}=\sum_{l=0}^{r}\genfrac{[}{]}{0pt}{}{N-1}{2l+1}
 \quad(r\geq2),
\]
with explicit checks through $N=8$ and order $\alpha'^7$. His  
permutation structures are organized by the inverse Solomon descent
algebra. 

Equation (3.24) of Ref.~\cite{Mafra:2021wok} nevertheless provides a useful constructive starting point for our coefficient blocks \(W_n^{(k)}\). In Mafra's notation, an \(N\)-point motivic open string component is decomposed according to
\begin{equation}
 A_{f_2^m f_M}(1,2,\ldots,N)
 =
 \sum_{\ell=1}^{2m+1}
 \ \sum_{\substack{
 P_1P_2\cdots P_\ell=23\cdots N\\
 P_i\neq\varnothing}}
 C^{f_2^m f_M}_{1|P_1,\ldots,P_\ell},
 \label{eq:mafra-deconcatenation}
\end{equation}
where the inner sum runs over the ordered deconcatenations of the word
\(23\cdots N\). The corresponding KK-like selection rules retain
\(\ell=1\) for \(A_{f_M}\), \(\ell=3\) for \(A_{f_2f_M}\), and
\(\ell=1,3,\ldots,2m+1\) for \(A_{f_2^m f_M}\) with \(m\geq2\).
For every pure-open amplitude appearing on the right-hand side of
eq.~\req{FINAL}, this decomposition can be applied after relabelling the word
\(23\cdots N\) to the relevant ordering of the open-string legs. At a
fixed total order \(\alpha'^k\), let
\(W_{n,\mathrm{cand}}^{(k)}\) denote the matrix obtained by collecting all
coefficient rows generated in this way, including the expansion of the
open--closed kernel \req{kernel}, the specialization to the symmetric collinear
configuration \req{leadingcoll} subject to \req{half}, the extraction of the strict \(\epsilon^0\) coefficient \req{solvefor},
and the subsequent reduction to the BCJ vector \(\mathbf C_n\). Our
coefficient block may then be characterized constructively as
\[
 W_n^{(k)}
 =
 \operatorname{RowBasis}
 \bigl(W_{n,\mathrm{cand}}^{(k)}\bigr),
 \qquad
 \operatorname{rk}W_n^{(k)}
 =
 \left[\begin{matrix}n-3\\ k\end{matrix}\right]
\]
in the cases verified here. Thus Mafra's deconcatenation formula provides
an overcomplete generating prescription for the rows of \(W_n^{(k)}\);
the open--closed projection, the special collinear kinematics and the
BCJ relations are responsible for the additional linear dependencies
which must be removed.
This should not yet be regarded as a closed all-multiplicity formula\footnote{We stress that direct substitution into eq.~\req{FINAL} uses the same
pure-open multiplicity, \(N=n\). The shifted identification
\(n=N+2\) employed below serves only to compare the numerical Stirling
entries through \(n-3=N-1\); it is not the multiplicity map entering the
open--closed relation in~\req{FINAL}.} for
\(W_n^{(k)}\). A general combinatorial translation of the higher-mass
objects \(C^{f_2^m f_M}_{1|P_1,\ldots,P_\ell}\) into SYM amplitudes and
Mandelstam invariants is not presently known. Such a translation is
available for the simplest cases \(C_{1|P}\) and
\(C^{f_2}_{1|P_1,P_2,P_3}\), the latter through the \(S\)-map construction.
Extending it to the higher components would make the above prescription
fully explicit.

An important difference is that the counting of
Ref.~\cite{Mafra:2021wok} is cumulative in the power $m$ of \(f_2\).
It gives the dimension of each individual motivic component
\(A_{f_2^m f_M}\), rather than a decomposition of one common amplitude
space into mutually complementary graded sectors. For example, at
\(N=6\) one obtains
\[
 d_{6,2}
 =
 \genfrac{[}{]}{0pt}{}{5}{1}
 +\genfrac{[}{]}{0pt}{}{5}{3}
 +\genfrac{[}{]}{0pt}{}{5}{5}
 =
 24+35+1
 =
 60 ,
\]
and
\[
 d_{6,3}
 =
 \genfrac{[}{]}{0pt}{}{5}{1}
 +\genfrac{[}{]}{0pt}{}{5}{3}
 +\genfrac{[}{]}{0pt}{}{5}{5}
 +\genfrac{[}{]}{0pt}{}{5}{7}
 =
 60 ,
\]
since \(\genfrac{[}{]}{0pt}{}{5}{7}=0\). Thus the same Stirling
components reappear in the basis dimensions of successive motivic
amplitudes.
By contrast, in our corresponding \(n=N+2=8\) open--closed collinear
problem, the same three numbers occur once as separate independent
blocks,
\[
 \operatorname{rk}W_8^{(1)}=24,
 \qquad
 \operatorname{rk}W_8^{(3)}=35,
 \qquad
 \operatorname{rk}W_8^{(5)}=1.
\]
Together they form the \(60\)-dimensional gravitational half of a
single BCJ-reduced reconstruction space. Our Stirling hierarchy is
therefore non-cumulative: each odd Stirling sector contributes {\it once} to
the complete gravitational basis, while each even sector contributes
{\it once} to the complementary space of homogeneous constraints.

In addition to this non-cumulative organization, our construction acts
on a different space and uses further physical input.  We first apply the mixed open--closed monodromy kernel
\req{kernel}, then specialize to \req{leadingcoll} with \req{half}, extract the pole-subtracted strict subleading collinear coefficient \req{solvefor}, and finally reduce the result to an $(n-3)!$-dimensional BCJ basis.  After
these steps, the even Stirling blocks $W_n^{(2j)}$ become homogeneous, kinematically
weighted constraints, while the odd blocks $W_n^{(2j+1)}$ become EYM and
higher-derivative EYM data.  Their combined matrix has full rank through
$n=9$ and yields the reconstruction formula
\req{reconstruction-formula}.  These features are not part of the
pure-open KK-like counting of Ref.~\cite{Mafra:2021wok}.  Our result is
therefore a more refined BCJ-reduced, open--closed collinear realization of a related
Stirling hierarchy, rather than a rederivation of the pure-open KK based result.

Ref.~\cite{Mafra:2021wok} also argues that the BCJ-preserving
$\zeta_M$ tails do not change the basis dimension within a fixed
$\zeta_2^r\zeta_M$ class.  This is consistent with our observation that
the odd-zeta matrices $M_{2s+1}$ do not produce additional projected rank
in the examples studied.  It does not, however, imply the open--closed
collinear rank decomposition \req{stirling-dual-decomposition}.

\subsection{A possible algebraic interpretation}

Successively stacking
\(W_n^{(1)},W_n^{(2)},\ldots,W_n^{(n-3)}\) produces a nested sequence
of row spaces. According to \req{stirling-dual-decomposition}, the
rank increases at the \(k\)-th step by
\(\genfrac{[}{]}{0pt}{}{n-3}{k}\). This nested rank hierarchy is all
that is meant here by the term Stirling filtration. Since
\(\genfrac{[}{]}{0pt}{}{n-3}{k}\) counts permutations of \(n-3\)
elements with \(k\) cycles, the result points toward the cycle and
descent structure of \(S_{n-3}\). The inverse Solomon descent algebra
in Ref.~\cite{Mafra:2021wok} is the closest established algebraic
antecedent of this counting.

The determinant conjecture \eqref{conjectural-Dn} makes this cycle
interpretation more concrete. For a fixed unordered bipartition
\(I|I^c\), the exponent
\[
 (|I|-1)!(|I^c|-1)!
\]
is  the number of permutations whose two cycles have supports
\(I\) and \(I^c\): there are \((|I|-1)!\) cyclic orderings on \(I\)
and \((|I^c|-1)!\) cyclic orderings on \(I^c\). Summing these multiplicities over all unordered bipartitions with
both blocks non-empty reproduces the degree formula
\req{degree-conjectural-Dn}, and hence gives
\[
\genfrac{[}{]}{0pt}{}{n-2}{2},
\]
the number of permutations of the \(n-2\) hard labels with two cycles.
This explains directly why the conjectural determinant has the degree
required by eq.~\eqref{degree-complete-determinant}. It also identifies
the two-block partition lattice of the hard labels as a natural
combinatorial structure behind its factors.

A distinct and interesting geometric analogy comes from the
type-\(A\) braid hyperplane arrangement.  In the present setting, the braid
arrangement consists of the collision hyperplanes
\[
 H_{ij}
 =
 \left\{
 (z_1,\ldots,z_{n-3})\in\mathbb C^{n-3}
 \ \middle|\
 z_i=z_j
 \right\},
 \qquad
 1\leq i<j\leq n-3.
\]
Their complement is the ordered configuration space:
\begin{equation}
 \operatorname{Conf}_{n-3}(\mathbb C)
 :=
 \mathbb C^{n-3}\setminus\bigcup_{i<j}H_{ij}
 =
 \left\{
 (z_1,\ldots,z_{n-3})\in\mathbb C^{n-3}
 \ \middle|\
 z_i\neq z_j\ \text{for}\ i\neq j
 \right\}\ .
 \label{conf}
\end{equation}
Permuting the \(n-3\) coordinates leaves the arrangement invariant.
Its reflection group is \(S_{n-3}\), conventionally called the Weyl
group of type \(A_{n-4}\).  This is the meaning of ``type \(A\)'' in
the present discussion.
The Poincar\'e polynomial of the configuration space \req{conf} is:
\begin{equation}
 P_{\operatorname{Conf}_{n-3}(\mathbb C)}(t)
 =
 \prod_{j=1}^{n-4}(1+jt)
 =
 \sum_{k=1}^{n-3}
 \genfrac{[}{]}{0pt}{}{n-3}{k}\,
 t^{n-3-k}\ .
 \label{Poincare}
\end{equation}
Consequently, the rank of the \(k\)-th block agrees numerically with
the corresponding Betti number,
\begin{equation}
 \operatorname{rk}W_n^{(k)}
 =
 \dim H^{\,n-3-k}
 \!\left(\operatorname{Conf}_{n-3}(\mathbb C)\right)
 =
 \genfrac{[}{]}{0pt}{}{n-3}{k}.
 \label{rank-cohomology-comparison}
\end{equation}

The Orlik--Solomon algebra describes the ordinary cohomology of this
hyperplane complement \cite{OrlikSolomon:1980}.  Its generators may be
represented by the logarithmic one-forms
\[
 \omega_{ij}=d\log(z_i-z_j)
\]
subject to the Arnold--Orlik--Solomon relations.  This description is
particularly natural for disk integrals, whose singular divisors arise
when marked points collide.

On the other hand, the Solomon descent algebra is a different object: it is a subalgebra
of the group algebra of \(S_{n-3}\), organized by descent classes of
permutations.  The relation between the two structures arises because
the same symmetric group \(S_{n-3}\) permutes the collision
hyperplanes and acts on their cohomology.  More specifically, the
Lehrer--Solomon decomposition of
\[
 H^q\!\left(\operatorname{Conf}_{n-3}(\mathbb C)\right)
\]
is organized by conjugacy classes of permutations with \(n-3-q\)
cycles \cite{LehrerSolomon:1986}.  Setting \(q=n-3-k\), the same
\(k\)-cycle statistic that determines
\(\genfrac{[}{]}{0pt}{}{n-3}{k}\) therefore labels the corresponding
cohomological degree.  More direct relations between components of
Solomon's descent algebra and the Orlik--Solomon algebra of Coxeter
arrangements were established in
\cite{DouglassPfeifferRoehrle}.

The configuration space \eqref{conf} also admits an exact relation to
the genus-zero moduli space $\mathcal M_{0,r}$ relevant for open string disk amplitudes.  For \(r\geq3\), the moduli space of \(r\)
ordered marked points on the Riemann sphere is
\begin{equation}
 \mathcal M_{0,r}
 :=
 \operatorname{Conf}_{r}(\mathbb P^1)
 \big/
 \operatorname{PGL}(2,\mathbb C)
 \simeq
 \operatorname{Conf}_{r-3}
 \!\left(\mathbb C\setminus\{0,1\}\right),
 \label{definition-M0N}
\end{equation}
where three marked points have been fixed at \(0,1,\infty\).
To relate this space to \(\operatorname{Conf}_{\ell}(\mathbb C)\),
choose two labels, say \(z_1\) and \(z_2\).  The variables \(z_1\) and
\(z_2-z_1\) describe the overall translation and non-zero scale,
respectively.  After normalizing \(z_1=0\), \(z_2=1\), and adjoining
the point at infinity, the remaining configuration defines an element
of \(\mathcal M_{0,\ell+1}\).  Thus, for \(\ell\geq2\),
\begin{equation}
 \operatorname{Conf}_{\ell}(\mathbb C)
 \simeq
 \mathbb C\times\mathbb C^\times
 \times\mathcal M_{0,\ell+1}.
 \label{conf-M0-relation}
\end{equation}
Specializing to \(\ell=n-3\) ($r=\ell+1=n-2$) gives
\begin{equation}
 \operatorname{Conf}_{n-3}(\mathbb C)
 \simeq
 \mathbb C\times\mathbb C^\times
 \times\mathcal M_{0,n-2},
 \qquad
 P_{\operatorname{Conf}_{n-3}(\mathbb C)}(t)
 =
 (1+t)\,
 P_{\mathcal M_{0,n-2}}(t).
 \label{conf-M0-specialization}
\end{equation}
At the level of ordinary cohomology, the only additional factor is
therefore the degree-one contribution \((1+t)\) from
\(\mathbb C^\times\), since \(\mathbb C\) is contractible.
The appearance of \(\mathcal M_{0,n-2}\) is striking in the present
context, since \(n-2\) is precisely the number of original open-string
insertions in the mixed open--closed disk amplitude.  This suggests
that the cohomological Stirling structure may primarily be associated
with the open-string ordering data, while the closed-string insertion
induces the mixed monodromy projection studied here.

These geometric and representation-theoretic correspondences do not
yet identify our kinematically weighted row spaces with the above
algebraic components.  In particular, we have not shown that the row
spaces generated by \(W_n^{(k)}\) are preserved by a natural
\(S_{n-3}\) action, nor have we constructed an intertwining map to
\[
H^{\,n-3-k}
\!\left(\operatorname{Conf}_{n-3}(\mathbb C)\right).
\]
The determinant conjecture sharpens the algebraic target by exposing
a bipartition and two-cycle structure, but it does not by itself
provide such a map.  We therefore regard the Orlik--Solomon comparison
and the relation to \(\mathcal M_{0,n-2}\) as a possible geometric
interpretation of the observed ranks, rather than as a derivation of
\req{stirling-dual-decomposition}.

\section{Universality and effective-field-theory descendants}
\label{sec:universality}

The derivation above was presented for amplitudes with purely gluonic external states. Its essential ingredients, however, are more general. We note that although the collinear limit~\eqref{eq: col kin} is parametrized in four dimensions and the open--closed string relations are formulated in ten dimensions, the final results, when expressed in terms of Lorentz products, can be applied in general dimensions. For an explicit realization and its relation to four-dimensional spinor-helicity variables, see~\cite{Jin}. Moreover, the open--closed momentum kernels depend only on the ordering and momenta of the external states, and not on their particle species, while collinear factorization is controlled locally by the two adjacent legs. The same
hierarchy therefore has natural extensions both to other components of a
supersymmetric gauge multiplet
\cite{StiebergerTaylor:2007SUSY,Sondergaard:2009Matter} and to effective
field theories obtained by transmutation operators or dimensional reduction
\cite{Cachazo:2014xea,Cheung:2017ems,Cheung:2017yef,Dong:2021qai,Dong:2024klq}.

\subsection{External fermions and scalars in \texorpdfstring{$\mathcal N=4$}{N=4} super--Yang--Mills theory}
The results of this paper can also be applied to super--Yang--Mills (SYM) theory with other external states. For example, let us consider $\mathcal N=4$ SYM theory in four dimensions.
Assign a component state to each of the $n-2$ hard legs,
\be\label{hard-state-assignment}
 \boldsymbol\Phi=(\Phi_1,\ldots,\Phi_{n-2}),
 \qquad
 \Phi_i\in\{g,\lambda^A,\bar\lambda_A,\phi^{AB}\},
 \qquad
 N_{\rm F}(\boldsymbol\Phi)\in2\mathbb N,
\ee
and define
\be\label{susy-component-amplitude}
 A_\rho^{\boldsymbol\Phi}(\epsilon):=
 A_{\mathcal N=4\,\mathrm{SYM}}
 \bigl(1^{\Phi_1},
 \rho(2^{\Phi_2},\ldots,(n-2)^{\Phi_{n-2}}),
 (n-1)^g,n^g\bigr),
 \qquad \rho\in S_{n-3}.
\ee
Here $g$ denotes a gluon of either helicity,
$\lambda^A$ and $\bar\lambda_A$, with $A=1,\ldots,4$, denote the gluinos
and anti-gluinos, and $\phi^{AB}=-\phi^{BA}$ denotes the six real scalar
states of the $\mathcal N=4$ multiplet, subject to the usual reality
condition.  The quantity $N_{\rm F}(\boldsymbol\Phi)$ counts all gluino and
anti-gluino external states.  It must be even for a non-vanishing component
amplitude.  Thus, any subset of the hard gluons may be replaced by other
states in the multiplet, provided the total number of external fermions is
even.  The permutation $\rho$ acts on the hard labels together with their
assigned states, whereas the two adjacent collinear particles remain
gluons.  Since neither the momentum kernel nor the
gluon--gluon splitting channel is changed by the spectator states, the
coefficient matrices $c_\rho^{(k)}$ in \req{eq:set1} are unchanged.  The
only replacement is at the level of the amplitudes,
\be\label{susy-replacement}
 A_{\rm YM}\ \longrightarrow\
 A_{\mathcal N=4\,\mathrm{SYM}}(\boldsymbol\Phi),
 \qquad
 A_{\rm EYM}^{(2j)}\ \longrightarrow\
 A_{{\rm EYM},\boldsymbol\Phi}^{(2j)} .
\ee
Here $A_{{\rm EYM},\boldsymbol\Phi}^{(0)}$ denotes the matter-coupled EYM component
amplitude with the same hard external states, and
$A_{{\rm EYM},\boldsymbol\Phi}^{(2j)}$ denotes the corresponding component of the
supersymmetric higher-derivative completion.  In particular, if the hard
configuration contains a non-zero even number of gluinos and anti-gluinos,
the amplitudes on the right-hand side of the odd Stirling sectors must
contain those same external fermions.  With this understanding, the
relations take the componentwise form
\be\label{susy-hierarchy}
\begin{aligned}
\ap^{2k-1}\sum_{\rho\in S_{n-3}}c_\rho^{(2k-1)}
 A_\rho^{\boldsymbol\Phi}(\epsilon)\Big|_{\epsilon^0}
 &=\frac{g^2}{\kappa x}\,
 A_{{\rm EYM},\boldsymbol\Phi}^{(2k-2)}(1,\ldots,n-2;P),\\
\ap^{2k}\sum_{\rho\in S_{n-3}}c_\rho^{(2k)}
 A_\rho^{\boldsymbol\Phi}(\epsilon)\Big|_{\epsilon^0}
 &=0 .
\end{aligned}
\ee
As in \req{eq:set1}, the finite coefficient is defined after transforming
the universal leading collinear subtraction.  This extension does not
apply without modification when one of the two collinear particles is a
fermion: in that case the splitting amplitude and the possible parent
state are different.

\subsection{Transmutation and dimensional reduction}

A broad web of amplitude relations can be generated by dimensional reduction
of CHY integrands, or equivalently by Lorentz-invariant differential
transmutation operators.  A particularly useful class consists of special
reductions under which higher-dimensional gluon polarizations become
lower-dimensional Goldstone modes
\cite{Cachazo:2014xea,Cheung:2017ems,Cheung:2017yef}.  These constructions
relate YM and gravity amplitudes to the non-linear sigma model (NLSM), Born--Infeld theory, the
special Galileon (sGal), and mixed theories.  Universal expansions into mixed
amplitudes and extensions to stringy pion models have been discussed in
Refs.~\cite{Dong:2021qai,Dong:2024klq}.

Let $\mathfrak T$ denote such a linear map and suppose that it is regular
in the collinear parameter.  Since $\mathfrak T$ acts on polarization or
state data whereas $c_\rho^{(k)}$ is a polynomial in Mandelstam invariants,
\be
 \mathfrak T\!\left(c_\rho^{(k)}A_\rho\right)
 =c_\rho^{(k)}\,\mathfrak T A_\rho,
 \qquad
 \mathfrak T\!\left(A_\rho\Big|_{\epsilon^m}\right)
 =\left(\mathfrak T A_\rho\right)\Big|_{\epsilon^m}.
\ee
It follows that the hierarchy \req{eq:set1} descends schematically to
\be\label{transmuted-hierarchy}
\begin{aligned}
\ap^{2k-1}\sum_\rho c_\rho^{(2k-1)}
 A_{\mathfrak T{\rm YM},\rho}\Big|_{\epsilon^0}
 &=\frac{g^2}{\kappa x}\,
 \mathfrak T A_{\rm EYM}^{(2k-2)},\\
\ap^{2k}\sum_\rho c_\rho^{(2k)}
 A_{\mathfrak T{\rm YM},\rho}\Big|_{\epsilon^0}
 &=0 .
\end{aligned}
\ee
The map must be applied to the complete collinear expansion, including the
leading residue, before imposing the pole-subtraction convention of
section~\ref{strictprojection}.  In this way the pole-mixing term is transformed
together with the amplitude and no additional finite contribution is lost.
The relations themselves therefore survive transmutation.  Their
interpretation as a complete reconstruction system also survives whenever
the reduced kinematics leave the matrix ${\cal M}_n$ at full rank.  Extra
selection rules may make some of the transmuted gravitational data vanish;
the NLSM provides an important example.

\paragraph{The NLSM example.}
To make the reduction explicit, choose two distinguished legs, which we
take to be $1$ and $2$, and first act with the trace operator
$\partial/\partial e_1\cdot e_2$.  For every remaining polarization, special dimensional reduction (DR) of
Refs.~\cite{Cachazo:2014xea,Dong:2024klq} is realized by the
$2d$-dimensional embedding
\be\label{DRI-embedding}
 p_a^M=(p_a^\mu,0),
 \qquad
 e_a^M=(0,i p_a^\mu),
 \qquad a\notin\{1,2\},
\ee
and therefore acts on Lorentz contractions as
\be\label{DRI-rules}
 \begin{aligned}
 \mathrm{DR}:\qquad
 e_a\cdot p_b&\ \longrightarrow\ 0,
 &e_a\cdot e_b&\ \longrightarrow\ -p_a\cdot p_b,
 &p_a\cdot p_b&\ \longrightarrow\ p_a\cdot p_b,\\[-1mm]
 &&& e_a,e_b\notin\{e_1,e_2\}.&&
 \end{aligned}
\ee
It is convenient to denote the complete operation by
\be
 \mathfrak D_{\rm I}^{(1,2)}
 :=\mathrm{DR}\circ\partial_{e_1\cdot e_2}.
\ee
After absorbing the conventional overall normalization into
$\mathfrak D_{\rm I}^{(1,2)}$, its action on a color-ordered YM amplitude is
\be\label{nlsm-reduction}
 \mathfrak D_{\rm I}^{(1,2)}A_{\rm YM}(1,2,\ldots,m)
 =A_{{\rm NLSM}\oplus\phi^3}
   (1_\phi,2_\phi,3,\ldots,m)
 =A_{\rm NLSM}(1,2,\ldots,m).
\ee
The two $\phi^3$ labels in the intermediate expression merely implement
the two distinguished rows of the reduced CHY Pfaffian; with  two
such states the mixed amplitude equals the pure NLSM partial amplitude.

Applying the same operation to the EYM side makes the vanishing of the
lowest odd sector manifest.  Indeed, a one-graviton EYM amplitude can be
expanded schematically as
\be\label{one-graviton-EYM-expansion}
 A_{\rm EYM}(1,\ldots,m;P)
 \ \propto\
 \sum_{\ell=1}^{m-1}
 (\varepsilon_P\cdot X_\ell)\,
 A_{\rm YM}(1,\ldots,\ell,P,\ell+1,\ldots,m),
 \qquad
 X_\ell:=\sum_{j=1}^{\ell}p_j,
\ee
where one of the two factorized graviton polarizations is denoted by
$\varepsilon_P$ \cite{Stieberger:2016lng}.  Since the graviton is not one
of the two distinguished legs, DR~I includes
$\varepsilon_P\cdot p_j\to0$ and hence
\be\label{DRI-EYM-zero}
 \mathfrak D_{\rm I}^{(1,2)}
 A_{\rm EYM}(1,\ldots,m;P)=0.
\ee
Equivalently, the open-string/YM states are mapped to pions while the
gravitational state is mapped to a special-Galileon state, so that
\be
 \mathfrak D_{\rm I}^{(1,2)}A_{\rm EYM}
 =A_{{\rm NLSM}\oplus{\rm sGal}}
   (1,\ldots,m;P_{\rm sGal})=0.
\ee
This mixed amplitude contains  one special-Galileon particle.  Its
vanishing is therefore also a direct consequence of the parity selection
rule of the mixed NLSM--special-Galileon theory, according to which an odd
number of special-Galileon states gives zero.

At the leading field theory limit, this proves the vanishing of the ordinary EYM
right-hand side directly.  The same conclusion applies to the
higher-derivative or stringy odd sectors whenever their mixed
NLSM--special-Galileon descendants preserve the odd-particle selection rule.  Under this condition, all strict $\epsilon^0$ data on
the right-hand side of the transmuted odd sectors vanish, while the even
sectors remain homogeneous.  Provided the reduced reconstruction matrix has
generic full rank,
\req{transmuted-hierarchy} then implies
\be\label{nlsm-coll-zero}
 A_{\rm NLSM}(1,\rho,n-1,n)\Big|_{\epsilon^0}=0,
 \qquad
 A_{\rm NLSM}(1,\rho,n-1,n)_{\rm coll.}
 =\epsilon\,C_{\rho,{\rm NLSM}}^{(1)}+O(\epsilon^2).
\ee
Thus an adjacent collinear limit of a tree-level NLSM amplitude starts at
order $\epsilon$, rather than at order $\epsilon^0$.  This agrees with the
direct CHY analysis of Ref.~\cite{Nandan:2016CHY}, where
the NLSM CHY integrand was found to have leading adjacent-collinear behavior
of order $\epsilon$.  Equation \req{nlsm-coll-zero} also illustrates why
the transformed hierarchy need not have non-zero data in every Stirling
sector: transmutation preserves the relations, but the target theory may
impose additional state-counting selection rules.

\section{Conclusions and further directions}

\subsection{Concluding remarks}

We have shown that the complete set of $(n-3)!$ strict subleading collinear
data \req{CollExpa} of YM amplitudes is naturally organized by open--closed string disk
amplitudes. The resulting structure is governed by unsigned Stirling numbers
of the first kind,
$
c(n-3,k)=\left[{n-3\atop k}\right].
$
The even sectors provide higher-order BCJ-like relations, while the odd sectors
provide a gravitational basis in terms of EYM amplitudes and their
higher-derivative corrections.

More explicitly, the even Stirling sectors satisfy
$
\sum_{j\geq 1}
\left[{n-3\atop 2j}\right]
=
\frac12 (n-3)!,
$
and reduce the original set of YM collinear data to a basis of dimension
$\frac12 (n-3)!$. The complementary odd sectors satisfy
$
\sum_{j\geq 0}
\left[{n-3\atop 2j+1}\right]
=
\frac12 (n-3)!,
$
and are represented by gravitational amplitudes. The first odd sector gives
the ordinary EYM amplitudes, while the higher odd sectors correspond to
$F^4$-, $F^6$-, and higher-derivative-corrected EYM amplitudes.
At $n=6$ the decomposition
$
6=2+3+1
$
corresponds to two independent EYM amplitudes, three quadratic BCJ-like relations and one
$F^4$-corrected EYM amplitude. At $n=7$ one obtains
$
24=6+11+6+1,
$
and at $n=8$
$
120=24+50+35+10+1.
$
Eventually, at \(n=9\) we find  the Stirling pattern
$
720=120+274+225+85+15+1,
$
including one additional sextic BCJ-like relation beyond the quadratic and
quartic relations found explicitly.
The pattern observed in the low-multiplicity examples is summarized in 
table \ref{Table2}.

\begin{table}[H]
\hskip-0.5cm
\renewcommand{\arraystretch}{1.2}
\begin{tabular}{|c|c|c|c|c|}
\hline
$n$ 
& strict YM data 
& independent EYM amplitudes
& BCJ-like relations
& gravitational basis
\\
\hline
$5$ & $2$   & $1$  & $1$  & $1$
\\
$6$ & $6$   & $2$  & $3$  & $3$
\\
$7$ & $24$  & $6$  & $12$ & $12$
\\
$8$ & $120$ & $24$ & $60$ & $60$
\\
$9$ & 720 & 120 & 360 & 360
\\
$n$ & $(n-3)!$ & $(n-4)!$ & $\frac12 (n-3)!$ & $\frac12 (n-3)!$
\\
\hline
\end{tabular}
\caption{
The strict subleading collinear sector contains $(n-3)!$ YM data. The BCJ-like
relations reduce this number by one half. The remaining half-dimensional space
is represented by EYM amplitudes and their higher-derivative corrections.
}\label{Table2}
\end{table}

\noindent
The central result is stronger than a counting statement.  After selecting independent rows in every Stirling block, the even and odd sectors form the square reconstruction matrix \req{even-odd-matrices}
\begin{equation}
 {\cal M}_n
 =
 \begin{pmatrix}
  {\cal R}_n\\[1mm]
  {\cal G}_n
 \end{pmatrix},
 \qquad
 \operatorname{rk}{\cal M}_n=(n-3)! ,
\end{equation}
at generic kinematics for all multiplicities verified through $n=9$.  Consequently, the pole-subtracted strict subleading coefficients are reconstructed from the gravitational data according to eq.~\req{reconstruction-formula}.  Thus the ordinary and higher-derivative EYM amplitudes are not merely additional constraints: they provide coordinates on the half-dimensional solution space selected by the homogeneous even-sector relations.

The explicit determinants at five  \req{det-five-point}, six \req{det-complete-six-point} and seven points \req{det-seven-point} provide a complementary test of this reconstruction.  Their generic non-vanishing establishes invertibility at these multiplicities and motivates the all-multiplicity factorization conjecture \req{conjectural-Dn}.  The factors associated with unordered bipartitions of the hard labels, together with the exponents $(|I|-1)!(|I^c|-1)!$, expose a two-cycle structure whose total degree is fixed independently by the Stirling decomposition.  The rank computations imply ${\cal D}_n\not=0$ through $n=9$, whereas the detailed factorization formula remains conjectural for $n\geq8$.

The Stirling organization is related to, but physically distinct from, the KK-like counting of motivic pure-open string amplitudes in Ref.~\cite{Mafra:2021wok}.  Mafra's basis dimensions are cumulative in successive powers of $\zeta_2$, whereas the present hierarchy is non-cumulative: every Stirling block contributes exactly once to a single BCJ-reduced reconstruction space.  Moreover, the mixed open--closed projection separates the blocks into homogeneous even-sector constraints and odd-sector gravitational data.  The inverse Solomon descent algebra provides the closest established algebraic antecedent of this counting.  The relation to the Orlik--Solomon algebra of the type-$A$ braid arrangement remains a possible geometric interpretation of the observed ranks rather than a derivation.

A complementary, and logically distinct, geometric direction is to construct the odd-sector gravitational data $A_{\rm EYM}^{(2j)}$ directly, rather than extracting them order by order from the mixed open--closed disk relation \req{FINAL}. Ordinary all-multiplicity EYM amplitudes admit both a CHY representation \cite{Cachazo:2014nsa} and a realization as intersection numbers of twisted differential forms on the moduli space of punctured Riemann spheres \cite{Mazloumi:2022lga}. It would be interesting to identify higher-derivative deformations of the corresponding CHY integrands or twisted cocycles that generate the complete tower
\[
A_{\rm EYM}^{(2j)}
\longleftrightarrow F^{2j+2}\text{-corrected EYM}\ ,
\qquad j\geq 0 .
\]
Such a construction could provide direct all-multiplicity expressions for the entries of the gravitational vector $A_n^{\rm grav}$ in eq.~\eqref{reconstruction-formula}, perhaps an independent derivation of the corresponding coefficient matrix $G_n$ and likewise the vector ${\boldsymbol C}_n$ comprising the subleading corrections. It might also render the observed Stirling ranks and determinant factorization manifest.

All results in the present work concern the symmetric energy splitting
\(x=\frac12\).  This choice is not merely a technical simplification:
the general pole-subtracted field-theory analysis shows that all
dependence of the strict subleading coefficients on the auxiliary
reference spinor \(r\) disappears at this point.  The reconstruction
matrix therefore acts on intrinsic collinear data.  Extending the
coefficient matrices and the reconstruction formula to arbitrary
\(x\) is the natural next step.  Away from the symmetric point the
reference-spinor dependence is generically restored and must be kept
as part of the off-collinear information.  Controlling this dependence
is also required for the Mellin transformation entering celestial
OPEs, as discussed in subsection \ref{celestial}.

\subsection{Mixed-helicity collinear pairs}\label{MixedHelicity}
Throughout this work, we have restricted the two collinear gluons to equal helicities,
\[
(h_{n-1},h_n)=(+,+)\qquad\text{or}\qquad(-,-).
\]
The corresponding closed-string state therefore has helicity \(+2\) or \(-2\), and the odd sectors of the reconstruction are represented by ordinary and higher-derivative EYM amplitudes with one external graviton. An important extension is obtained by considering the mixed-helicity configurations \((+,-)\) and \((-,+)\). As observed already in Ref.~\cite{Stieberger:2014hba}, these configurations represent the two scalar polarizations of the closed-string state, conventionally described as a complex dilaton and its conjugate. Equivalently, one may introduce states of definite left--right parity,
\[
{\cal A}_{\phi}
 =\frac{1}{\sqrt{2}}\left({\cal A}_{+-}+{\cal A}_{-+}\right),
\qquad
{\cal A}_{a}
 =\frac{1}{i\sqrt{2}}\left({\cal A}_{+-}-{\cal A}_{-+}\right),
\]
corresponding, up to conventions, to the dilaton and axionic scalar.

At the symmetric point \(x=\frac12\), however, the two left--right parity sectors behave differently.  Denoting the corresponding strict coefficients 
\req{eq:strict-projection-main} by
\[
 C_{\phi,\rho}^{(0)}
 =
 \frac{1}{\sqrt{2}}
 \left(
 C_{+-,\rho}^{(0)}+C_{-+,\rho}^{(0)}
 \right),
 \qquad
 C_{a,\rho}^{(0)}
 =
 \frac{1}{i\sqrt{2}}
 \left(
 C_{+-,\rho}^{(0)}-C_{-+,\rho}^{(0)}
 \right),
\]
the field-theory collinear expansion \cite{Jin} shows that the dependence on the reference spinor \(r\) cancels in the symmetric coefficient \(C_{\phi,\rho}^{(0)}\), whereas it generically survives in \(C_{a,\rho}^{(0)}\).  Thus the symmetric dilaton sector defines intrinsic data on the strict collinear surface, 
in direct analogy with the equal-helicity coefficients in \req{Situation}.  The antisymmetric axionic sector, by contrast, retains information about the direction from which the collinear surface is approached and therefore does not belong to the same reference-independent reconstruction space.

This distinction is reminiscent of the invariant vertex operators constructed for closed-string scattering on the real projective plane in Ref.~\cite{Bischof:2025rwx}.  There, invariance under the crosscap involution is obtained by combining a left--right vertex insertion with its image under the world-sheet identification.  That construction is specific to the 
\(\mathbb R {\mathbb P}^2\) involution and does not by itself remove the reference-spinor dependence encountered here, since \(r\) parametrizes an off-collinear kinematic direction rather than a world-sheet redundancy.  Nevertheless, it illustrates why combinations of definite left--right parity are the natural closed-string variables.
\(\mathbb R \mathbb P^2\) involution and does not by itself remove the reference-spinor dependence encountered here, since \(r\) parametrizes an off-collinear kinematic direction rather than a world-sheet redundancy.  Nevertheless, it illustrates why combinations of definite left--right parity are the natural closed-string variables.

At the level of the open--closed reduction, 
the monodromy kernels \req{kernel} and the coefficient system encoded in
\req{eq:set1}  are independent of the external helicities.  It is therefore natural to expect that, after componentwise subtraction of the two possible parent-helicity poles, the symmetric mixed-helicity coefficients can be reconstructed in terms of dilaton--Yang--Mills amplitudes and their higher-derivative corrections by an analogue of the hierarchy derived above.  We do not make the corresponding claim for the antisymmetric sector.  Its treatment would require either an enlarged reconstruction space retaining the \(r\)-dependent transverse data or an additional invariant completion, possibly formulated in terms of the gauge-invariant \(B\)-field strength and its four-dimensional axion dual.  We leave this question for future work.

\subsection{Further directions toward local celestial OPEs}\label{celestial}

Celestial operator-product coefficients are determined by collinear limits of momentum-space amplitudes \cite{Fan:2019emx,Fotopoulos:2019vac}. The leading singular OPE coefficients of gluons and gravitons in EYM were analyzed in Ref.~\cite{Fotopoulos:2019vac,PateRaclariuStromingerYuan2019,Fotopoulos:2020bqj}. Worldsheet constructions subsequently generated the singular OPE together with infinite towers of descendants \cite{AdamoBuCasaliSharma2021}, while all regular contributions are known in the MHV sector from twistor-string theory and inverse-soft recursion \cite{AdamoBuCasaliSharma2022,RenSchreiberSharmaWang2023}. These all-order MHV expressions are naturally organized in terms of soft-current descendants. Although this gives a closed representation of the OPE, it is not manifestly decomposed into ordinary single-point celestial operators, since individual descendant expressions retain kinematic data associated with both collinear legs.

The gravitational representation found here suggests a complementary organization. An EYM amplitude contains a single external graviton with momentum $P=p_{n-1}+p_n$. After Mellin transformation, it can therefore be interpreted as a matrix element of a gravitational insertion at the fused celestial point. More generally, we may define candidate local gravitational blocks ${\mathbb G}^{(2j)}_{\Delta_P,+2}(z_P,\bar z_P)$ operationally through
\begin{equation}
\left\langle
{\mathbb G}^{(2j)}_{\Delta_P,+2}(z_P,\bar z_P)
\prod_{a=1}^{n-2}
{\cal O}_{\Delta_a,J_a}(z_a,\bar z_a)
\right\rangle
\equiv
\widetilde A_{\rm EYM}^{(2j)}
\bigl(1,\ldots,n-2;P^{+2}\bigr),
\end{equation}
where $\widetilde A_{\rm EYM}^{(2j)}$ denotes the celestial transform of the ordinary or higher-derivative EYM amplitude entering \req{eq:set1}. In this interpretation, the odd Stirling sectors    \req{block-system}
provide independent gravitational matrix-element structures 
$W_n^{(2j+1)}$, whereas the even sectors impose homogeneous relations among the corresponding celestial OPE data.

Establishing this interpretation requires extending the present construction from the symmetric splitting $x=\frac12$ to arbitrary energy fraction $x$, followed by the appropriate Mellin transform. Although the strict coefficients studied here are reference-spinor
independent at \(x=\frac12\), their extension to arbitrary \(x\) will
generically reintroduce such dependence.  It will then be necessary to
determine how this off-collinear information is reorganized into
ordinary celestial descendants and to demonstrate that the resulting
OPE coefficients are independent of the spectator insertions. If these conditions are satisfied, the EYM representation would provide a manifestly single-point and gravitationally organized description of regular celestial OPEs beyond the MHV sector. The conformally soft limits of these gravitational blocks would furthermore determine whether the regular collinear data merely complete the local celestial OPE outside the wedge or induce genuine extensions or deformations of the celestial $Lw_{1+\infty}$ and ${\cal S}$ algebras.

\medskip\medskip
\goodbreak

\paragraph{Acknowledgements.}
We would like to thank Nima Arkani-Hamed,  Carolina Figueiredo, and Tomasz Taylor  for useful discussions.
This work is supported by the DFG grant 508889767 {\it 
Forschungsgruppe ``Modern foundations of scattering amplitudes''}. 


\appendix

\break
\section{World-sheet interpretation of pole mixing}
\label{app:pole-mixing}

The origin of the pole mixing term ${\cal B}$ is most transparent before the world-sheet integrals are expanded.  
A generic contribution to the disk correlator \req{FINAL} with $n-2$ open-string
vertices on the boundary and one closed-string vertex in the bulk has
the form \cite{Stieberger:2015kia}
\begin{align}
 F_n={}&V_{\rm CKG}^{-1}\,
 \delta^{(D)}\!\left(\sum_{i=1}^{n-2}p_i+q_1+q_2\right)
 \int\prod_{i=1}^{n-2}dx_i
 \prod_{1\leq r<s\leq n-2}
 |x_r-x_s|^{2\alpha'p_r\cdot p_s}(x_r-x_s)^{n_{rs}}
 \nonumber\\
 &\times\int_{\mathbb H_+}d^2z\,
 (z-\bar z)^{2\alpha'q_1\cdot q_2+n_0}
 \prod_{i=1}^{n-2}
 (x_i-z)^{2\alpha'p_i\cdot q_1+n_i}
 (x_i-\bar z)^{2\alpha'p_i\cdot q_2+\bar n_i}.
 \label{eq:mixed-disk-integral-main}
\end{align}
Here $V_{\rm CKG}$ is the volume of the conformal Killing group.  The
integers $n_{rs},n_0,n_i$ and $\bar n_i$, as well as the tensor
coefficients which are suppressed in \req{eq:mixed-disk-integral-main}, are fixed
by the vertex-operator correlator.  Their detailed values are not required  
for the argument; what matters is that the noninteger powers encode the
Koba--Nielsen factors and their branch phases.

In the contour deformation of the complex world-sheet disk integral leading  to the open-string representation
\req{FINAL}, one writes $z=x+iy$ and $\bar z=x-iy$ and deforms the
$y$ contour according to the prescription of
Ref.~\cite{Stieberger:2015kia}.  On the deformed contour, the
combinations
\begin{equation}
 \xi:=x+iy,
 \qquad
 \eta:=x-iy
 \label{eq:xi-eta-boundary-coordinates-main}
\end{equation}
are integrated as independent real variables.  They  effectively become the positions of the two boundary insertions carrying the left- and
right-moving momenta $q_1$ and $q_2$, respectively. Together with the original punctures $x_i$ the ordering of
$\xi$, $\eta$ specifies a particular
open-string integration chamber and its associated monodromy phase.
Throughout this appendix, ${\cal I}_\rho$ denotes the chamber integral after
stripping off the overall chamber-dependent monodromy/contour weight.  At the level of a fixed chamber this is the piecewise-constant branch phase, together with its contour orientation; after chambers are collected, these weights produce the sine and exponential factors displayed in \req{FINAL}.  Apart from this external weight, no position-dependent factor is removed from ${\cal I}_\rho$: it contains the complete Koba--Nielsen and
OPE integrand over the chamber.  The stripped weight is restored when the
chambers are assembled into the full disk relation.  In particular, its
off-collinear expansion is not discarded: its first variation contributes
to $c_{\rho,1}$ and hence to the pole-mixing term ${\cal B}$.
The change of variables has Jacobian
$\partial(x,y)/\partial(\xi,\eta)=i/2$, and before  further chamber
decomposing its real domain is
$-\infty<\xi<\infty$ and $\xi<\eta<\infty$
\cite{Stieberger:2015kia}.
Within a chamber with $\eta>\xi$, introduce
\begin{equation}
 t:=\eta-\xi>0,
 \qquad
 u:=\frac{\xi+\eta}{2}.
 \label{eq:collision-coordinates-main}
\end{equation}
The degeneration $z\to\bar z$ is then the boundary collision
$t\to0$, while $u$ is the collision point along the boundary.  The
endpoint analysis is local: choose a fixed $t_0>0$ smaller than the
distance to any other boundary of the integration chamber and split
the integral into $0<t<t_0$ and its complement.  The complementary
region is analytic at the massless locus.  If $\mathcal I_\rho$ denotes
the monodromy-weight-stripped but otherwise complete integral of chamber
$\rho$, this separation is
schematically
\begin{equation}
 \mathcal I_\rho
 =I_\rho^{\rm end}+I_\rho^{\rm rem},
 \label{eq:full-chamber-endpoint-split-main}
\end{equation}
where $I_\rho^{\rm rem}$ contains the complementary integration region
and any other terms analytic in this channel.  Thus
$I_\rho^{\rm end}$ is not the full chamber integral: it denotes only
the local contribution associated with the massless collision
endpoint.  Moreover, the notation ``end'' does not mean that a factor
has been subtracted from the full integrand and discarded.  Rather, we
have restricted the full chamber integral to a small neighbourhood of
this endpoint and will factor out its universal singular dependence;
the rest of the chamber remains in $I_\rho^{\rm rem}$.  Accordingly,
eq.~\eqref{eq:full-chamber-endpoint-split-main} is an additive decomposition
of the chamber integral, not a decomposition of the monodromy weight
multiplying it.
In the endpoint region, the rescaling $t=t_0\tau$ maps the integration
range to $0<\tau<1$; the resulting factor $t_0^{\alpha's}$ and the
integrations over $u$ and the spectator punctures can be absorbed into
a smooth function.  This rescaling is the reason that the canonical
local integral below runs from $0$ to $1$: it is a normalized
coordinate on an arbitrarily small endpoint neighbourhood, not a claim
that the original separation $\eta-\xi$ has unit range.  Relabeling
$\tau$ as $t$, the contribution of chamber $\rho$ containing the
massless channel is locally of the form
\begin{equation}
 I_\rho^{\rm end}(s,\epsilon)
 =\int_0^1dt\,t^{\alpha's-1}\phi_\rho(t,s,\epsilon),
 \qquad s:=2q_1\cdot q_2=s_{n-1,n},
 \label{eq:local-endpoint-integral-main}
\end{equation}
where the factor $t^{\alpha's}$ descends from the Koba--Nielsen factor
$(z-\bar z)^{2\alpha'q_1\cdot q_2}$, while the additional $t^{-1}$ is
the massless term in the boundary OPE.  All remaining OPE coefficients
and spectator factors are collected in $\phi_\rho$, which is smooth in
$t$ at the endpoint and analytic in $s$ near $s=0$.  In other words,
the universal collision factor $t^{\alpha's-1}$ has been factored out,
while the integrations over $u$ and the spectator punctures, together
with the remaining OPE data, have been packaged into $\phi_\rho$.
Nothing is discarded in making this definition.  The factors coupling
the bulk insertion to a boundary puncture do not generate an additional
singularity in this isolated collision region.

For $\operatorname{Re}(\alpha's)>0$, adding and subtracting the endpoint value
in \req{eq:local-endpoint-integral-main} gives
\begin{equation}
 \begin{aligned}
 I_\rho^{\rm end}(s,\epsilon)
 &=\frac{\phi_\rho(0,s,\epsilon)}{\alpha's}
   +I_\rho^{\rm end,reg}(s,\epsilon),\\
 I_\rho^{\rm end,reg}(s,\epsilon)
 &:=\int_0^1dt\,t^{\alpha's-1}
 \bigl[\phi_\rho(t,s,\epsilon)
       -\phi_\rho(0,s,\epsilon)\bigr].
 \end{aligned}
 \label{eq:endpoint-subtraction-main}
\end{equation}
The first term contains the complete massless factorization pole.  The
second is regular because smoothness implies
\begin{equation}
 \phi_\rho(t,s,\epsilon)-\phi_\rho(0,s,\epsilon)
 =t\,\psi_\rho(t,s,\epsilon),
 \label{eq:endpoint-regularity-main}
\end{equation}
with $\psi_\rho$ regular at $t=0$.  It therefore becomes
$\int_0^1dt\,t^{\alpha's}\psi_\rho$ and has a finite limit at $s=0$.
To identify the two pieces relevant for the subsequent strict
projection, following \req{MandelO} impose the collinear scaling
$s=\sigma(r)\epsilon^2$, with
$\sigma(r):=2\langle Pr\rangle[Pr]$, and expand the endpoint data as
\begin{equation}
 \begin{aligned}
 \phi_\rho(0,\sigma\epsilon^2,\epsilon)
 &=\epsilon R_{\rho,1}
   +\epsilon^2R_{\rho,2}+O(\epsilon^3),\\
 I_\rho^{\rm end,reg}(\sigma\epsilon^2,\epsilon)
 &=I_{\rho,0}^{\rm end,reg}+O(\epsilon).
 \end{aligned}
 \label{eq:endpoint-collinear-expansion-main}
\end{equation}
The absence of an ${\cal O}(\epsilon^0)$ term follows from the fact that the physical Yang–Mills collinear singularity is only ${\cal O}(\epsilon^{-1})$: since $1/s\sim\epsilon^{-2}$, the residue of the massless endpoint pole must vanish at least linearly in $\epsilon$.
At a generic momentum partition, $R_{\rho,2}$ need not vanish.  There
is, however, a simplification for the equal-helicity channel on the
symmetric $x=1/2$ path considered below.  With a left--right symmetric
closed-string state and a symmetric momentum-conservation completion,
$\epsilon\to-\epsilon$ exchanges the two collinear legs.  Their color-ordered three-point OPE coefficient is odd under this exchange, whereas the fused lower-point correlator is even: it depends on the
combined momentum $P_{\rm tot}=P+\epsilon^2r$ and on the endpoint
spectator factors through symmetric combinations such as
$s_{i,n-1}+s_{i,n}$.  Consequently,
\begin{equation}
 \phi_\rho(0,\sigma\epsilon^2,-\epsilon)
 =-\phi_\rho(0,\sigma\epsilon^2,\epsilon)\ 
 \Longrightarrow \ 
 R_{\rho,2}=0,
 \label{eq:symmetric-endpoint-residue-main}
\end{equation}
and the endpoint residue contains only odd powers of $\epsilon$. 

In the symmetric equal-helicity setup, eq.
\req{eq:endpoint-subtraction-main} therefore becomes
\begin{equation}
 I_\rho^{\rm end}(\sigma\epsilon^2,\epsilon)
 =
 \underbrace{\frac{1}{\epsilon}
 \frac{R_{\rho,1}}{\alpha'\sigma}}_{\rm leading\ pole}
 +
 \underbrace{I_{\rho,0}^{\rm end,reg}}_{
 \rm retained\ finite\ endpoint\ contribution}
 +O(\epsilon).
 \label{eq:endpoint-kept-pieces-main}
\end{equation}
Thus the regular integral $I_{\rho,0}^{\rm end,reg}$ is retained.  Away
from this symmetric setup, a nonzero
$R_{\rho,2}/(\alpha'\sigma)$ would likewise belong to the retained
finite endpoint contribution; it would not be part of the mixing term
defined below.  The leading $1/\epsilon$ term is not itself part of the
finite coefficient.  After the reduction \req{EXPA}, its coefficient is
represented by $C_\rho^{(-1)}$, with the string dressing absorbed into $c_\rho$.  Only its product with the first variation
$c_{\rho,1}$ contributes at order $\epsilon^0$ and is omitted:
\begin{equation}
 \underbrace{c_{\rho,1}C_\rho^{(-1)}}_{
 \rm omitted\ pole\text{-}mixing}
 \simeq\mathcal B_{\rho}.
 \label{eq:omitted-endpoint-mixing-main}
\end{equation}
Accordingly, the first term of
\req{eq:endpoint-subtraction-main} is not removed as a whole; only this
finite mixing with its leading pole is excluded by the strict
projection.

This refines the origin of the mixing term.  The first
off-collinear variation of the original monodromy row and the first variation of the period matrix $F$ both contribute to $c_{\rho,1}$.  Their product
with the universal YM pole produces
$\mathcal B= \sum_\rho
c_{\rho,1}C_\rho^{(-1)}$.  Thus
$\mathcal B$ contains no new hard YM collinear coefficient:
it is the variation of the effective string prefactor multiplying the
already known leading YM factorization pole.

\end{document}